\documentclass[sigconf,screen]{acmart}

\usepackage{enumitem}
\usepackage{doi}
\usepackage{balance}

\AtBeginDocument{
  }

\setcopyright{cc}
\setcctype{by}
\acmDOI{10.1145/3832783.3834377}
\acmYear{2026}
\copyrightyear{2026}
\acmISBN{979-8-4007-2882-2/2026/10}
\acmConference[ASE '26]{Proceedings of the 41st IEEE/ACM International Conference on Automated Software Engineering}{October 12--16, 2026}{Munich, Germany}
\acmBooktitle{Proceedings of the 41st IEEE/ACM International Conference on Automated Software Engineering (ASE '26), October 12--16, 2026, Munich, Germany}
\acmSubmissionID{ase26main-p1071-p}
\received{2026-03-26}
\received[accepted]{2026-06-18}

\newif\ifdiff
\difffalse

\ifdiff
  \usepackage{xcolor}
  \newcommand{\added}[1]{\textcolor{blue}{#1}}
\else
  \newcommand{\added}[1]{#1}
\fi

\begin{document}

\title{Programming by Chat: A Large-Scale Behavioral Analysis of 11,579 Real-World AI-Assisted IDE Sessions}

\author{Ningzhi Tang}
\authornote{Both authors contributed equally to this research.}
\orcid{0000-0002-8555-2351}
\affiliation{
  \institution{University of Notre Dame}
  \city{Notre Dame}
  \state{IN}
  \country{USA}
}
\email{ntang@nd.edu}

\author{Chaoran Chen}
\authornotemark[1]
\orcid{0000-0002-9161-4088}
\affiliation{
  \institution{University of Notre Dame}
  \city{Notre Dame}
  \state{IN}
  \country{USA}
}
\email{cchen25@nd.edu}

\author{Zihan Fang}
\orcid{0009-0009-2151-2922}
\affiliation{
  \institution{Vanderbilt University}
  \city{Nashville}
  \state{TN}
  \country{USA}
}
\email{zihan.fang@vanderbilt.edu}

\author{Gelei Xu}
\orcid{0009-0003-6854-1487}
\affiliation{
  \institution{University of Notre Dame}
  \city{Notre Dame}
  \state{IN}
  \country{USA}
}
\email{gxu4@nd.edu}

\author{Maria Dhakal}
\orcid{0009-0004-0428-3164}
\affiliation{
  \institution{University of Notre Dame}
  \city{Notre Dame}
  \state{IN}
  \country{USA}
}
\email{mdhakal@nd.edu}

\author{Yiyu Shi}
\orcid{0000-0002-6788-9823}
\affiliation{
  \institution{University of Notre Dame}
  \city{Notre Dame}
  \state{IN}
  \country{USA}
}
\email{yshi4@nd.edu}

\author{Collin McMillan}
\orcid{0009-0005-0887-1083}
\affiliation{
  \institution{University of Notre Dame}
  \city{Notre Dame}
  \state{IN}
  \country{USA}
}
\email{cmc@nd.edu}

\author{Yu Huang}
\orcid{0000-0003-2730-5077}
\affiliation{
  \institution{Vanderbilt University}
  \city{Nashville}
  \state{TN}
  \country{USA}
}
\email{yu.huang@vanderbilt.edu}

\author{Toby Jia-Jun Li}
\orcid{0000-0001-7902-7625}
\affiliation{
  \institution{University of Notre Dame}
  \city{Notre Dame}
  \state{IN}
  \country{USA}
}
\email{toby.j.li@nd.edu}

\begin{abstract}
  IDE-integrated AI coding assistants, which operate conversationally within developers' working codebases with access to project context and multi-file editing, are rapidly reshaping software development. However, empirical investigation of this shift remains limited: existing studies largely rely on small-scale, controlled settings or analyze general-purpose chatbots rather than codebase-aware IDE workflows. We present, to the best of our knowledge, the first large-scale study of real-world conversational programming in IDE-native settings, analyzing 74,998 developer messages from 11,579 chat sessions across 1,300 repositories and 899 developers using Cursor and GitHub Copilot. These chats were committed to public repositories as part of routine development, capturing in-the-wild behavior. Our findings reveal three shifts in how programming work is organized: conversational programming operates as progressive specification, with developers iteratively refining outputs rather than specifying complete tasks upfront; developers redistribute cognitive work to AI, delegating diagnosis, comprehension, and validation rather than engaging with code and outputs directly; and developers actively manage the collaboration, externalizing plans into persistent artifacts, and negotiating AI autonomy through context injection and behavioral constraints. These results provide foundational empirical insights into AI-assisted development and offer implications for the design of future programming environments.
\end{abstract}



\maketitle

\section{Introduction}

IDE-integrated AI assistants such as GitHub Copilot (Chat)\footnote{\url{https://github.com/features/copilot}} and Cursor\footnote{\url{https://cursor.com/}} have evolved beyond autocomplete into multi-turn conversational systems. Unlike general-purpose chat interfaces (e.g., ChatGPT in the browser), these assistants operate directly within the developer's working codebase, with access to project context, coordinated multi-file editing, terminal execution, and runtime inspection. This emerging practice, which we refer to as \textit{AI-assisted conversational programming in IDE-native settings}, represents a shift in how developers work: rather than writing code directly, developers increasingly express intent and iteratively steer AI-generated outputs through dialogue. In a more extreme form, this delegation can resemble \textit{``vibe coding''}~\cite{sarkar2025vibe,chou2025building}, where substantial implementation is offloaded to AI with limited inspection; existing reports indicate that some codebases are now 95\% AI-generated\footnote{\url{https://techcrunch.com/2025/03/06/a-quarter-of-startups-in-ycs-current-cohort-have-codebases-that-are-almost-entirely-ai-generated/}}.

This shift indicates that IDE-native conversational programming is rapidly becoming a prominent mode of software development. Yet, despite widespread adoption, empirical understanding of how developers actually converse with these assistants in real-world environments remains limited. Characterizing these interactions is essential both for understanding modern development workflows and for designing more effective next-generation assistants. Currently, the empirical evidence is constrained in two complementary ways.\looseness=-1

\textbf{First,} existing studies lack either scale or ecological validity.
Video-based studies analyze publicly shared YouTube recordings in which developers code while narrating their process to an audience~\cite{sarkar2025vibe,chou2025building}, but this performative setting may systematically alter behavior through audience effects~\cite{alaboudi2019exploratory}, and the largest prompt-level analysis of this kind covers only 254 prompts~\cite{chou2025building}. 
Controlled user studies assign developers researcher-defined tasks on unfamiliar codebases~\cite{kumar2025ai,chen2025code,geng2025exploring}, a setting that differs markedly from everyday development, where goals are self-directed, context accumulates over time, and interaction is embedded in ongoing work~\cite{sjoberg2002conducting}. As a result, the range of observable goals and interaction styles remains constrained.
Finally, interview- and practitioner-based accounts~\cite{pimenova2025good,aiskillformation2026} are also subject to recall and self-presentation bias~\cite{lethbridge2005studying,goffman2023presentation}. Together, these studies provide valuable early insight, but they cannot capture how developers interact with AI coding assistants across the diverse self-directed tasks, codebases, and user populations of everyday software development.

\textbf{Second,} large-scale studies of developer-LLM interactions have primarily focused on browser-based chat rather than IDE-integrated environments. For example, DevGPT~\cite{xiao2024devgpt} curated developer-shared ChatGPT conversation links from GitHub, enabling subsequent studies of usage patterns~\cite{li2025unveiling}, code quality~\cite{siddiq2024quality}, and refactoring practices~\cite{alomar2024refactor}. More recently, CodeChat~\cite{zhong2025developer} assembled developer-LLM conversations from WildChat~\cite{zhao2024wildchat} and analyzed conversation-level topics and code defects. These datasets have enabled important large-scale analyses, but they do not capture interaction in IDE-native settings where assistants operate with repository context, tool access, and direct editing capabilities within an ongoing development workflow rather than as isolated query-answer exchanges.

\begin{figure}[t]
\centering
\includegraphics[width=\linewidth]{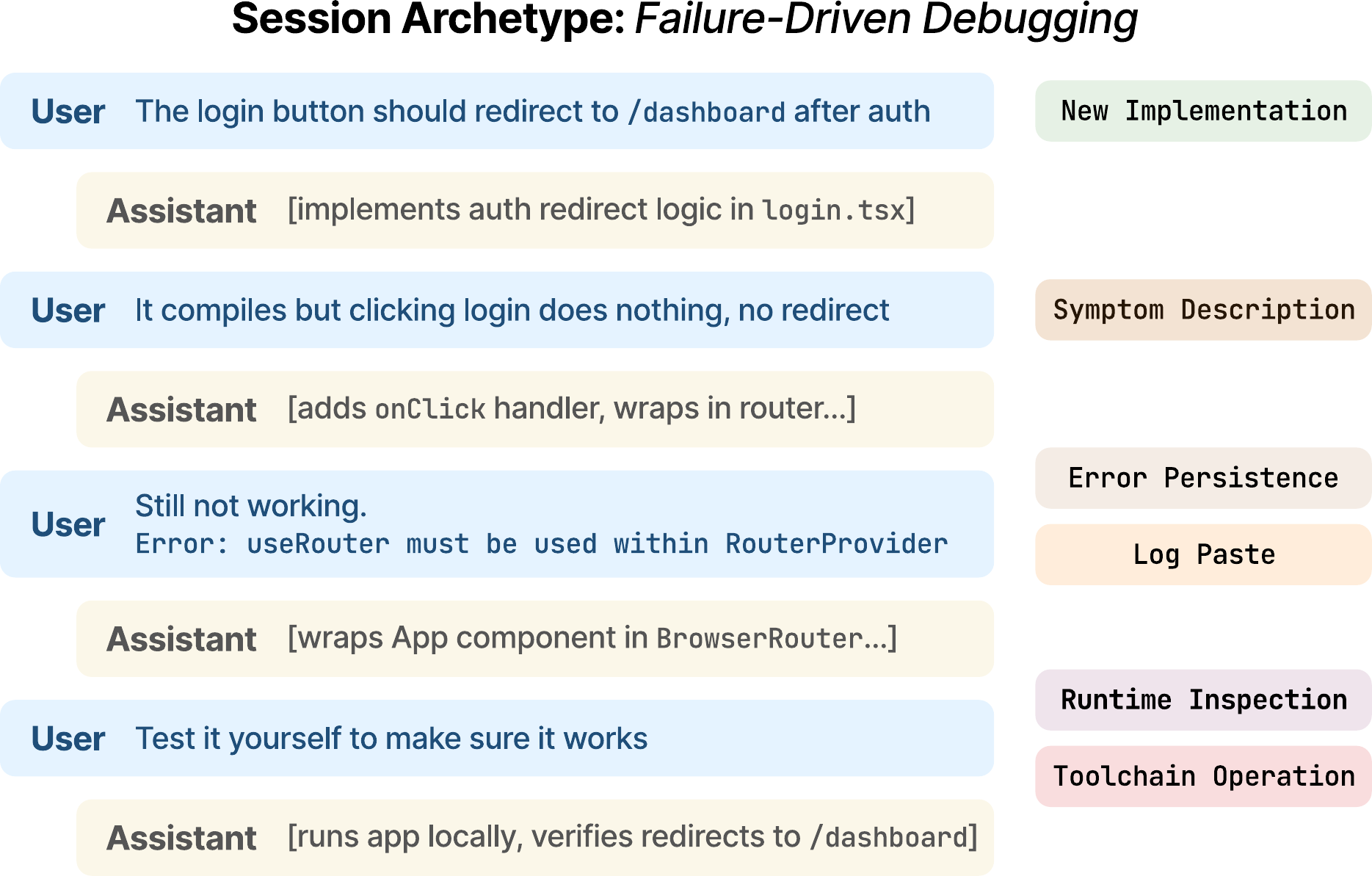}
\caption{Illustrative conversational programming session annotated with behavioral intent labels. The interaction exemplifies the \textit{Failure-Driven Debugging} archetype.}
\label{fig:illustrative_session}
\end{figure}

To address these limitations, we present, to the best of our knowledge, the first large-scale empirical study of AI-assisted conversational programming in IDE-native settings. Our dataset comprises 74,998 developer messages from 11,579 chat sessions across 1,300 repositories and 899 developers using Cursor and GitHub Copilot. These sessions were neither required nor observed by researchers. Developers committed their chat histories to public repositories as part of their workflow, capturing self-directed development activity in the wild. The dataset reflects substantial diversity: tasks range from web development to AI/ML engineering, messages are written in over 20 different natural languages, and repositories span a wide range of scales and maturity levels. This large-scale behavioral evidence enables us to characterize how developers structure tasks, delegate work to AI assistants, and recover from failures in IDE-native conversational programming. Figure~\ref{fig:illustrative_session} illustrates a representative chat session.

Using this dataset, we characterize conversational programming at two complementary levels, developer message intents and session-level structure, investigating the following research questions (RQs):

\begin{itemize}[leftmargin=2em]
    \item \textbf{RQ1:} \textit{What behavioral intents do developers express in their messages to AI coding assistants?} We develop a multi-label behavioral intent taxonomy through iterative abductive coding~\cite{timmermans2012theory}, comprising 7 main categories and 20 subcategories. We then use an LLM-based classifier to label all 74,998 messages, validated against a human-annotated reference set of 400 messages (macro-averaged F1 = 0.802).
    \item \textbf{RQ2:} \textit{What recurring behavioral archetypes and intent dynamics characterize conversational programming sessions?} We represent each session as an ordered sequence of intent labels and apply hierarchy-aware edit-distance-based~\cite{levenshtein1966binary} sequence clustering~\cite{kaufman2009finding} to 4,864 sessions with at least four messages, identifying six recurring behavioral archetypes. We then analyze intent dynamics through within-session transition patterns, session-boundary transitions, and session evolution from opening turns to later messages.
\end{itemize}

We highlight the following key results.
\begin{enumerate}[leftmargin=2em]
    \item \textbf{Conversational programming operates as progressive specification rather than upfront task description.} Developers steer the AI toward desired outputs through successive instructions (Finding~1); opening turns frame the task before sessions shift toward responding to AI outputs and emerging failures (Finding~11); and iterative modification and debugging tend to persist across consecutive turns (Finding~9).
    \item \textbf{Developers redistribute cognitive work to AI.} They report symptoms and machine outputs rather than diagnosing bugs (Finding 2), inquire about system behavior rather than reading code (Finding 3), and delegate validation to AI through both static code review and runtime inspection requests (Finding~4).
    \item \textbf{Developers actively manage the collaboration rather than treating AI as a passive tool.} They request persistent planning documents to externalize intent (Finding~5), negotiate AI autonomy through context injection and explicit action constraints (Finding~6), and start new chat sessions to refresh context while preserving task continuity (Finding~10).
    \item \textbf{Conversational programming exhibits structured interaction modes at the session level.} Most sessions are short and task-focused, but a long-tailed minority of sessions sustains extended iterative refinement (Finding~7). Six recurring archetypes emerge from clustering, ranging from failure-driven debugging to extended co-development (Finding~8).
\end{enumerate}
\section{Background and Related Work}

\subsection{Empirical Studies of AI Programming in IDE}

AI coding assistants have evolved from autocomplete tools into conversational and increasingly agentic systems. Early systems (e.g., the initial release of GitHub Copilot in 2021) supported only inline code completion, offering localized assistance within a single file or completion context, without invoking external tools, executing code, or coordinating edits across files. Empirical studies from this period~\cite{barke2023grounded,tang2024developer,liang2024large,nam2024using,tang2025exploring,tang2025naturaledit} reflect these limitations, examining interactions that were largely single-turn and narrowly scoped, in which the assistant primarily served as a reactive suggestion engine for short code spans within the developer's active cursor context. Sustained multi-turn delegation and iterative steering, which are central to more capable assistants, therefore remained largely beyond the scope of this earlier body of work.

This began to change with the rise of conversational interaction in IDEs. Tools such as Cursor and later versions of GitHub Copilot moved beyond inline completion to support chat-based interaction within the IDE and across broader project contexts. By February 2025, this shift had become culturally visible: Karpathy coined the term \textit{``vibe coding''} to describe accepting AI-generated code without reading diffs and pasting errors without diagnosis\footnote{\url{https://x.com/karpathy/status/1886192184808149383}}.

This emerging practice has attracted growing empirical attention through several forms of evidence. Video- and observation-based studies analyze publicly shared sessions, including YouTube and Twitch livestreams, to characterize prompting, delegation, and debugging behavior~\cite{sarkar2025vibe,chou2025building}. For example, Chou~\emph{et al.}~\cite{chou2025building} analyze livestreams and opinion videos, showing that vibe coding ranges from near-total delegation to more inspection-heavy collaboration. A smaller body of work draws on interviews, online discourse, and practitioner accounts to identify recurring themes such as co-creation, flow, trust, and concerns about quality assurance~\cite{pimenova2025good,fawzy2025vibe}. Controlled and field studies have examined increasingly agentic workflows. For example, Chen~\emph{et al.}~\cite{chen2025code} compare GitHub Copilot with the CLI-based agent OpenHands and find that higher automation can reduce user effort while raising transparency concerns. Kumar~\emph{et al.}~\cite{kumar2025ai} further show that incremental collaboration often outperforms one-shot delegation. Together, these studies provide important early insights, but they remain largely small-scale, observed, retrospective, or task-bounded, leaving in-the-wild IDE-native conversational programming undercharacterized.

\subsection{Large-Scale Studies of Coding with LLMs}

Existing empirical data for AI-assisted coding span benchmarks that evaluate model capabilities under standardized conditions~\cite{jimenez2023swe,zhao2025vibe,zhong2025vibe,deng2025swe,fang2025dpo,chi2025edit}, artifact-level datasets that capture development outcomes (e.g., pull request acceptance)~\cite{li2025rise}, \added{and field measurements of AI tool usage at scale~\cite{sergeyuk2026evolving,cui2026effects}}. While valuable, these resources measure task performance or downstream products rather than the interaction process itself. A separate line of work examines developer-provided context artifacts for coding agents, such as Cursor rules and \texttt{CLAUDE.md} files~\cite{jiang2025beyond,mohsenimofidi2025context}, but these capture persistent project-level specifications rather than dynamic multi-turn interaction.
The closest line of work to ours examines developer-LLM conversations directly, but primarily in general-purpose chatbots rather than IDE-native settings. DevGPT~\cite{xiao2024devgpt} and CodeChat~\cite{zhong2025developer} curate developer conversations from browser-based interfaces, i.e., ChatGPT and WildChat~\cite{zhao2024wildchat}, enabling follow-up analyses~\cite{siddiq2024quality,alomar2024refactor,li2025unveiling,hao2024empirical,fang2025comparative}. 
Crucially, none of these datasets capture interaction in IDE-native settings, where assistants operate directly with repository context and tool access as part of an ongoing development workflow. \added{This distinction is reflected in our findings: we replicate the predominantly short, task-delegation-oriented conversations observed in browser-based studies~\cite{hao2024empirical, li2025unveiling}, while behaviors such as runtime inspection and persistent planning documents surface only in IDE-native interactions (e.g., Findings 4--6).}

Our analytical approach draws on methodological precedents from large-scale studies of general-purpose LLM usage. Studies of ChatGPT, Claude, and Microsoft Copilot employ LLM-based classifiers to categorize conversations by topic, task, and usage context at scale~\cite{chatterji2025people,handa2025economic,aubakirova2026state,costa2025s}. These studies provide broad accounts of LLM use, but typically at the level of conversation topics or occupational tasks. Closer to our session-level focus, Mysore~\emph{et al.}~\cite{mysore2025prototypical} classify follow-up utterances in LLM-assisted writing and identify prototypical collaboration behaviors that account for much of the variation across multi-turn sessions. Related work operationalizes interaction dynamics such as autonomy and user oversight in agentic systems\footnote{\url{https://anthropic.com/research/measuring-agent-autonomy}}. These studies establish the feasibility of large-scale behavioral log analysis for characterizing human-LLM interaction. Our work applies and extends this approach to IDE-native conversational programming, a setting whose affordances and workflows remain underrepresented in prior analyses.
\section{Methodology}

We conducted a mixed-methods study of AI coding assistant chats from public GitHub repositories. For RQ1, we analyzed message-level intents using a taxonomy and LLM-based classification (Section~\ref{sec:method_rq1}). For RQ2, we analyzed session-level archetypes and intent dynamics (Section~\ref{sec:method_rq2}).

\subsection{Data Collection and Preparation}
\label{sec:data_collection}

We collected AI coding assistant chat histories from publicly available GitHub repositories via SpecStory\footnote{\url{https://specstory.com/}}, a developer tool that automatically exports chat histories as timestamped Markdown files stored in each project repository. SpecStory supports multiple AI coding tools, including IDE-integrated assistants (i.e., Cursor, GitHub Copilot) and CLI-based agents (e.g., Claude Code).

To retrieve these exported histories at scale, we queried the GitHub Code Search API\footnote{\url{https://docs.github.com/en/rest/search/search}} over \texttt{.specstory/history/}. Because the API caps results at 1,000 per query, we partitioned searches by date prefix (e.g., files starting with \texttt{2025-06-15}), leveraging SpecStory's date-prefixed filename convention. 
All data were collected on March 4, 2026, capturing sessions from September 2024 to March 2026, spanning SpecStory's availability and the early adoption of chat-based AI coding assistants in software development.

For each discovered file, we retrieved its raw Markdown via the Git Blobs API and parsed it into structured records. Specifically, we identified conversational turns using the role-prefixed markers introduced by SpecStory (e.g., ``User,'' ``Assistant,'' and ``Agent'') and treated the content associated with each marker as a single turn-level message for analysis. We provide the full parser implementation and format specifications in the replication package.
We manually inspected a random sample of 100 parsed sessions and found turn identification and role assignment to be accurate in all cases. We then retained only user messages for subsequent analysis, as they capture developer intents.

\paragraph{Data Cleaning and Exclusion.}

Each file is uniquely identified by its content hash (SHA), which we used as the deduplication key across search results. \added{This removed 3,512 duplicates. Manual inspection traced them mainly to developers duplicating chat histories within a single repository as snapshot backups (one developer alone accounted for 58\% of duplicates; 2,474 $\rightarrow$ 444 sessions after deduplication); and less to cross-repository copying via cloning, templating, or forking.} Our initial collection included 1,705 sessions exported from CLI-based agents (primarily Claude Code). We excluded these for two reasons: (1) their logs interleave user messages with tool outputs and system callbacks, obscuring user behavior; (2) they represent a different interaction paradigm in which users issue high-level tasks and agents autonomously plan and execute multi-step actions, rather than engaging in turn-by-turn conversational collaboration~\cite{chen2025code}. We therefore focused on IDE-integrated sessions and left CLI-agent behavior to future work.

After removing duplicates and excluding CLI-agent sessions, the dataset contained 76,231 user messages over 11,655 sessions. After the classification stage described in Section~\ref{sec:llm_classification}, we excluded 1,233 messages that do not have any classifiable behavioral intent, along with sessions with no remaining classifiable messages, yielding the final analytic dataset of \textbf{74,998 user messages} across \textbf{11,579 sessions}, from \textbf{1,300 unique repositories}, involving \textbf{899 developers}, identified by unique Git commit authors of chat history files.\looseness=-1

\paragraph{Dataset Characteristics.} 

The repositories in our dataset skewed toward personal and early-stage projects rather than established open-source software, with 67.3\% having a single contributor, consistent with the overall distribution of public GitHub repositories~\cite{kalliamvakou2014promises}. The primary languages were TypeScript (25.8\%), Python (22.0\%), JavaScript (13.1\%), and HTML (9.4\%), with a long tail of additional languages (e.g., Shell, Java, Rust, C++), each under 2.1\%. Most (90.8\%) were created in 2025 or later, with a median of 15 commits over 4 active days, although a long tail includes larger projects with thousands of commits and repositories dating back to 2013.

Automated language identification~\cite{lui2012langid} indicated highly multilingual user messages: English comprised 59.7\%, followed by Chinese (18.5\%) and Japanese (8.3\%), with the remainder across a long tail of other languages. As many prompts intermix natural language with code and technical tokens, these proportions should be treated as contextual indicators rather than precise linguistic annotations.

\subsection{Behavioral Intent Characterization}
\label{sec:method_rq1}

\subsubsection{Iterative Codebook Development}

We developed the \textit{behavioral intent} taxonomy through four rounds of iterative abductive coding~\cite{timmermans2012theory}. Prior schemes such as CUPS~\cite{mozannar2024reading} were designed for code-completion interactions and do not transfer well to multi-turn conversational dynamics, so we built the taxonomy from scratch. Drawing on Speech Act Theory~\cite{austin1975things,searle1969speech}, we labeled each message by its \textit{illocutionary act}, i.e., the intentional action performed (e.g., requesting, asserting, or directing), rather than by surface form or topic. We used a \textit{multi-label} scheme because many developers' turns in AI coding assistant chats contain multiple communicative acts. For example, \textit{``@config.yaml --- set up the database connection''} both provides context and requests implementation.

Three researchers with expertise in SE, AI, and HCI conducted the coding. In Round~1, we randomly sampled 300 messages with full conversational context and coded them through group discussion, resolving disagreements by consensus to establish an initial codebook. In Rounds~2--4, we used the same LLM-based classification pipeline later employed for full-dataset annotation (Section~\ref{sec:llm_classification}), but with the then-current version of the codebook, to generate \emph{provisional} labels for all messages. Under each main category, we also included a temporary \textsc{Others} subcategory for messages that did not fit existing definitions. These provisional labels were used to support diagnostic sampling for codebook refinement. Specifically, from each subcategory, including \textsc{Others}, we randomly sampled 40 messages for human review with conversational context. When boundary cases in a subcategory remained unresolved after initial discussion, we reviewed an additional 20 messages from that subcategory. We used these reviews to assess whether category definitions were sufficiently precise, whether boundary cases were handled consistently, and whether messages assigned to \textsc{Others} exhibited coherent recurring patterns warranting new subcategories. Across all rounds, coding decisions were grounded in the iterative interplay between empirical data patterns and our emerging theoretical framework, following abductive coding practice~\cite{timmermans2012theory}.

We assessed structural saturation by tracking the number of subcategories added or substantively revised after each round, following previous saturation criteria~\cite{lyu2025my,humbatova2020taxonomy}. The counts decreased monotonically across the four rounds (9, 5, 1, 0), with no new subcategories in the final round, indicating that the codebook had stabilized. The final taxonomy comprises \textbf{7 main categories} and \textbf{20 subcategories} (Table~\ref{tab:taxonomy}). The full codebook, including definitions, key signals, examples, and boundary notes, is provided in our replication package.

\begin{table*}[t]
\centering
\caption{Behavioral Intent Taxonomy with distribution statistics. \textit{\% Msg} denotes the percentage of messages assigned to each category or subcategory, calculated over the final analytic dataset of 74,998 messages. Due to multi-labeling, percentages sum to more than 100\% across categories, and subcategory percentages do not add up to their parent category total. \textit{\% Multi} denotes the percentage of messages carrying at least one additional label, computed within the same level.}
\label{tab:taxonomy}
\small
\begin{tabular}{l l r r}
    \toprule
    \textbf{Category / Subcategory} & \textbf{Definition} & \textbf{\% Msg} & \textbf{\% Multi} \\
    \midrule

    \textbf{\textsc{1. Code Authoring}} & \textbf{Requests to produce, modify, or adjust code.} & \textbf{34.53} & \textbf{35.71} \\
    \quad \textsc{1.1 New Implementation} & \quad Build something that does not yet exist as functional code. & 5.86 & 35.32 \\
    \quad \textsc{1.2 Iterative Modification} & \quad Adjust, constrain or refine existing or in-progress code. & 24.84 & 43.06 \\
    \quad \textsc{1.3 Alignment Correction} & \quad Redirect the AI when its output runs but misses the user's actual intent. & 7.21 & 83.63 \\
    \midrule

    \textbf{\textsc{2. Failure Reporting}} & \textbf{Report something broken, erroring, or behaving unexpectedly.} & \textbf{24.00} & \textbf{27.80} \\
    \quad \textsc{2.1 Log Paste} & \quad Paste machine-generated error logs or stack traces. & 8.84 & 39.74 \\
    \quad \textsc{2.2 Symptom Description} & \quad Describe a malfunction in natural language. & 14.77 & 50.76 \\
    \quad \textsc{2.3 Error Persistence} & \quad Signal that a previous fix attempt has failed. & 3.76 & 57.43 \\
    \midrule

    \textbf{\textsc{3. Inquiry}} & \textbf{Seek information, understanding, or advice without requesting code.} & \textbf{19.17} & \textbf{26.31} \\
    \quad \textsc{3.1 Planning \& Consultation} & \quad Plan next steps or seek advice on architectural choices. & 7.81 & 37.48 \\
    \quad \textsc{3.2 Project Comprehension} & \quad Understand the current state of the project or existing code. & 8.19 & 29.38 \\
    \quad \textsc{3.3 General Knowledge Query} & \quad Ask a project-agnostic technical or domain question. & 3.56 & 11.35 \\
    \midrule

    \textbf{\textsc{4. Context Specification}} & \textbf{Provide information or instructions to shape AI understanding or action rules.} & \textbf{14.08} & \textbf{65.42} \\
    \quad \textsc{4.1 Information Injection} & \quad Supply raw factual information for the AI to internalize. & 8.46 & 61.86 \\
    \quad \textsc{4.2 Behavior Specification} & \quad Negotiate AI autonomy, operating boundaries, or persona. & 6.14 & 77.88 \\
    \midrule

    \textbf{\textsc{5. Validation}} & \textbf{Ask the AI to evaluate or inspect code or execution output.} & \textbf{3.99} & \textbf{56.14} \\
    \quad \textsc{5.1 Code Review} & \quad Inspect code for correctness, quality, or security without running it. & 2.74 & 51.14 \\
    \quad \textsc{5.2 Runtime Inspection} & \quad Validate behavior by examining runtime output or test results. & 1.26 & 67.69 \\
    \midrule

    \textbf{\textsc{6. Delegation}} & \textbf{Instruct the AI to \added{perform development-workflow actions beyond authoring code.}} & \textbf{16.48} & \textbf{45.94} \\
    \quad \textsc{6.1 Documentation} & \quad Produce written artifacts such as docs, logs, or planning files. & 6.85 & 55.93 \\
    \quad \textsc{6.2 Toolchain Operation} & \quad Take action in the development environment via commands or tools. & 10.50 & 49.99 \\
    \midrule

    \textbf{\textsc{7. Workflow Control}} & \textbf{Manage session pace, direction, or state without new technical content.} & \textbf{11.47} & \textbf{24.36} \\
    \quad \textsc{7.1 Confirmation} & \quad Signal that the AI's output was correct or a step succeeded (e.g., ``works''). & 2.65 & 21.75 \\
    \quad \textsc{7.2 Continuation} & \quad Instruct the AI to proceed without introducing new requirements (e.g., ``continue''). & 5.54 & 16.67 \\
    \quad \textsc{7.3 Deferred Debugging} & \quad Terse fix command relying entirely on session context (e.g., ``fix it''). & 0.74 & 13.64 \\
    \quad \textsc{7.4 Deferred Implementation} & \quad Terse implementation command relying entirely on session context (e.g., ``edit it''). & 0.78 & 23.00 \\
    \quad \textsc{7.5 Sentiment Expression} & \quad Express an emotional or social state (e.g., greetings, frustration, gratitude). & 2.05 & 77.72 \\
    \bottomrule
\end{tabular}
\end{table*}

\subsubsection{LLM-Based Classification}
\label{sec:llm_classification}

We classified all 76,231 user messages with GPT-5 mini\footnote{\url{https://developers.openai.com/api/docs/models/gpt-5-mini}} via the OpenAI Batch API, with the finalized behavioral intent codebook provided as the system prompt. \added{We chose GPT-5 mini for its balance of classification capability and cost at our scale, following a comparable large-scale classification study~\cite{chatterji2025people}.}
We adopted an LLM-based classification approach because behavioral intent labeling is multi-label, context-dependent, and semantically nuanced. Prior work~\cite{wang2023large} suggests that LLMs can achieve strong performance on text classification tasks when guided by structured instruction and schema constraints.
Each classification request included the current message together with one turn of prior conversational context: the immediately preceding user message and the AI's most recent response. Context messages were truncated to 1,000 characters using a head-tail strategy (60\% prefix, 40\% suffix), and the current message was truncated to 10,000 characters. This context design captures the most immediate conversational context for intent disambiguation while keeping prompts at a consistent and tractable length, consistent with prior work on LLM-based classification of conversational data~\cite{chatterji2025people,mysore2025prototypical}. We enforced structured output via a JSON Schema, requiring the model to return a set of intent labels along with a brief textual justification for each. \added{We enabled the model's built-in reasoning mode (i.e., chain-of-thought~\cite{wei2022chain}), which produces intermediate reasoning steps before emitting the final JSON labels. This particularly helps disambiguate multi-intent and borderline cases.}

To validate classifier accuracy, we drew a stratified random sample of 400 messages (20 per subcategory) covering all 20 subcategories; because messages carry multiple labels, a single message may appear in more than one subcategory sample. Two researchers independently annotated all 400 messages using the finalized codebook and full session context. Inter-rater reliability was assessed per subcategory by treating each label as a binary present/absent decision and computing Cohen's $\kappa$ separately for each of the 20 subcategories, then macro-averaging; this yielded $\kappa = 0.669$, indicating substantial agreement~\cite{landis1977measurement}. Disagreements were resolved through discussion to produce an adjudicated gold-standard reference set. We then evaluated the LLM classifier against this reference set using the same per-category binary scheme, obtaining a macro-averaged F1 of 0.802 (precision = 0.774, recall = 0.851). The per-category results are reported in the replication package.

After classification, 1,233 messages (1.62\%) were assigned only to the residual \textsc{Others} category. Manual inspection confirmed that this category captures inputs without recoverable communicative intent, e.g., accidental keystrokes, parser artifacts such as \texttt{-{}-{}-}. We excluded these messages from all subsequent analyses and removed sessions with no classifiable messages, yielding the final analytic dataset described earlier.

\subsubsection{Inductive Qualitative Coding}

To characterize behavioral patterns, we conducted an inductive qualitative analysis following established thematic coding practices~\cite{lazar2017research}. For each of the 20 subcategories, we randomly sampled 80 messages. Because this analysis was interpretive in purpose, focusing on identifying recurring behavioral patterns rather than fixed categorical labels, we prioritized consensus-based thematic synthesis~\cite{mcdonald2019reliability}. To ensure rigor, two researchers independently reviewed each sampled message with full session context and conducted open coding to identify recurring behavioral patterns without predefined themes; they then discussed and reconciled their interpretations to produce consolidated thematic summaries. This sample size provided sufficient coverage for recurring themes to stabilize during analysis, consistent with prior discussions of thematic saturation in qualitative research~\cite{guest2006many}. \added{The resulting thematic summaries and illustrative quotes are reported alongside each finding in Section~\ref{sec:results_rq1}.}

\subsection{Session Archetypes and Dynamics Analysis}
\label{sec:method_rq2}

\subsubsection{Session Clustering}
\label{sec:clustering}

To complement the message-level analysis in Section~\ref{sec:method_rq1}, we examined whether AI-assisted conversational programming exhibits recurring session-level archetypes. Rather than clustering raw message text, we represented each session as a sequence of intent labels, thereby focusing on behavioral structure rather than surface linguistic variation. We restricted clustering to sessions with at least four user messages ($n = 4{,}864$), because shorter sessions contain too few within-session transitions to support meaningful sequence comparison using edit distance.

\paragraph{Session Representation.}

Each session $s$ is represented as an ordered sequence of label sets:
$$
s = (\ell_1, \ell_2, \ldots, \ell_T), \quad \ell_t \subseteq \mathcal{C}
$$
where $T$ is the number of user messages and $\mathcal{C}$ is the set of 20 subcategories in Table~\ref{tab:taxonomy}. Each $\ell_t$ is a non-empty set of intent labels ($|\ell_t| \geq 1$), naturally accommodating our multi-label scheme.

\paragraph{Hierarchy-Aware Edit Distance.}

We measure the dissimilarity $d(s, s')$ between two sessions $s = (\ell_1, \ldots, \ell_n)$ and $s' = (\ell_1', \ldots, \ell_m')$ using a weighted Levenshtein edit distance~\cite{levenshtein1966binary}, defined as the minimum total cost of transforming one sequence into the other via insertions, deletions, and substitutions, with operation costs:
\begin{itemize}
  \item \textit{Insertion or deletion} of an element $\ell_t$: cost $w_{\text{indel}} = 1.0$
  \item \textit{Substitution} of $\ell$ with $\ell'$: cost $\operatorname{sub}(\ell, \ell')$
\end{itemize}

The substitution cost between two label sets is defined as the mean pairwise label cost:
$$
\operatorname{sub}(\ell, \ell') = \begin{cases} 0 & \text{if } \ell = \ell' \\ \displaystyle\frac{1}{|\ell||\ell'|} \sum_{a \in \ell} \sum_{b \in \ell'} c(a, b) & \text{otherwise} \end{cases}
$$
where the label-level cost for subcategory labels $a, b \in \mathcal{C}$ is:
$$
c(a, b) = \begin{cases} 0 & \text{if } a = b \\ w_{\text{same}} = 0.5 & \text{if } \operatorname{main}(a) = \operatorname{main}(b),\ a \neq b \\ w_{\text{cross}} = 1.0 & \text{if } \operatorname{main}(a) \neq \operatorname{main}(b) \end{cases}
$$
where $\operatorname{main}(\cdot)$ maps a subcategory to its parent main category.

This three-tier cost structure encodes the taxonomy hierarchy: substitutions within a main category (e.g., \textsc{Symptom Description} $\to$ \textsc{Log Paste}) are penalized less than substitutions across categories (e.g., \textsc{Iterative Modification} $\to$ \textsc{Toolchain Operation}), following established practice in sequence analysis that substitution costs should decline as elements become more semantically similar~\cite{abbott2000sequence,studer2016matters}. When $|\ell| = |\ell'| = 1$, $\operatorname{sub}(\ell, \ell')$ reduces to $c(a, b)$, recovering the standard single-label case.

To prevent sessions of different lengths from being dominated by the cumulative cost of indel \added{(insertion and deletion)} operations, we normalized the raw edit distance by the longer sequence length:
$$
d(s, s') = \frac{d_{\text{raw}}(s, s')}{\max(|s|, |s'|)}
$$
where $d_\text{raw}(s, s')$ is the raw weighted edit distance computed by dynamic programming. Since $w_{\text{cross}} = w_{\text{indel}} = 1.0$, $d_{\text{raw}}(s, s') \leq \max(|s|, |s'|)$, so $d(s, s') \in [0, 1]$.

\paragraph{K-Medoids Clustering.}

We computed the full pairwise distance matrix over all $4{,}864$ sessions and applied K-Medoids (PAM)~\cite{kaufman2009finding} with k-medoids++ initialization~\cite{arthur2006k}, operating directly on the precomputed distance matrix and yielding actual sessions as cluster representatives (medoids). We evaluated $k \in \{2, \ldots, 20\}$ using the silhouette score~\cite{rousseeuw1987silhouettes}, which compares within-cluster similarity against neighboring clusters. The score peaked at $k = 6$ (silhouette $= 0.0703$), which we adopted as the final model. Absolute silhouette values were low across all tested $k$ ($0.0367$--$0.0703$), indicating weakly separated clusters. This likely reflects the nature of edit-distance sequence clustering, where sessions may share overlapping subsequences without forming sharply separated groups. We therefore interpret the six clusters conservatively as \textit{prototypical archetypes}, i.e., descriptive anchors of dominant behavioral patterns, rather than mutually exclusive natural classes. To characterize each archetype, two researchers independently examined the medoid sequence and a random sample of 30 sessions per cluster, then consolidated observations into a cluster label and narrative description. \looseness=-1
\added{We additionally created an interactive visualization tool displaying the t-SNE projection~\cite{van2008visualizing} with per-session label sequences and cluster assignments\footnote{Available at \url{https://conversational-programming-clusters.netlify.app/}}.}

\subsubsection{Intent Dynamics Analysis}
\label{sec:method_dynamics}

To complement clustering, we conducted three auxiliary analyses of behavioral intent dynamics.

For \textit{within-session transition analysis}, we computed lift-weighted Markov transition probabilities between consecutive intent labels within sessions. Lift is defined as $\text{lift}(i \to j) = P(j \mid i) / P(j)$, where $i$ and $j$ are intent labels; this normalizes raw transition counts by the marginal frequency of the target label, with lift $> 1$ indicating reinforced transitions and lift $< 1$ indicating suppressed ones. For multi-label messages, each consecutive pair was distributed over the Cartesian product of label sets with weight $1/(|\ell_t| \times |\ell_{t+1}|)$. To characterize short-range continuity, we also measured lengths of consecutive repetitions of the same subcategory within a session.

For \textit{session-boundary transition analysis}, we paired the last message of one session with the first message of the chronologically next session in the same repository and computed boundary lift, defined as $\text{lift}_{\text{boundary}}(i \to j) = P(\text{first}_j \mid \text{last}_i)/P(\text{first}_j)$, using the same multi-label weighting scheme as above. Comparing this quantity with within-session lift reveals whether session-boundary transitions differ from ordinary turn-to-turn transitions.

For \textit{session evolution analysis}, we compared opening messages with later turns in terms of intent-category proportions. We also examined how intent-category proportions shift across normalized message positions by mapping each session to a 100-position uniform grid via linear interpolation and fitting linear regressions to assess monotonic trends. Because opening turns strongly shaped these trajectories, we estimated an opening-excluded version that removed the first user message; unless otherwise noted, results refer to this specification. We applied the same procedure to message length (in characters) and labels per message.
\section{Results}

We report results in two parts. We first characterize the behavioral intent landscape at the message level (RQ1), and then examine session-level archetypes and dynamics (RQ2).

\subsection{RQ1. The Behavioral Intent Landscape}
\label{sec:results_rq1}

\begin{figure}[t]
\centering
\includegraphics[width=\columnwidth]{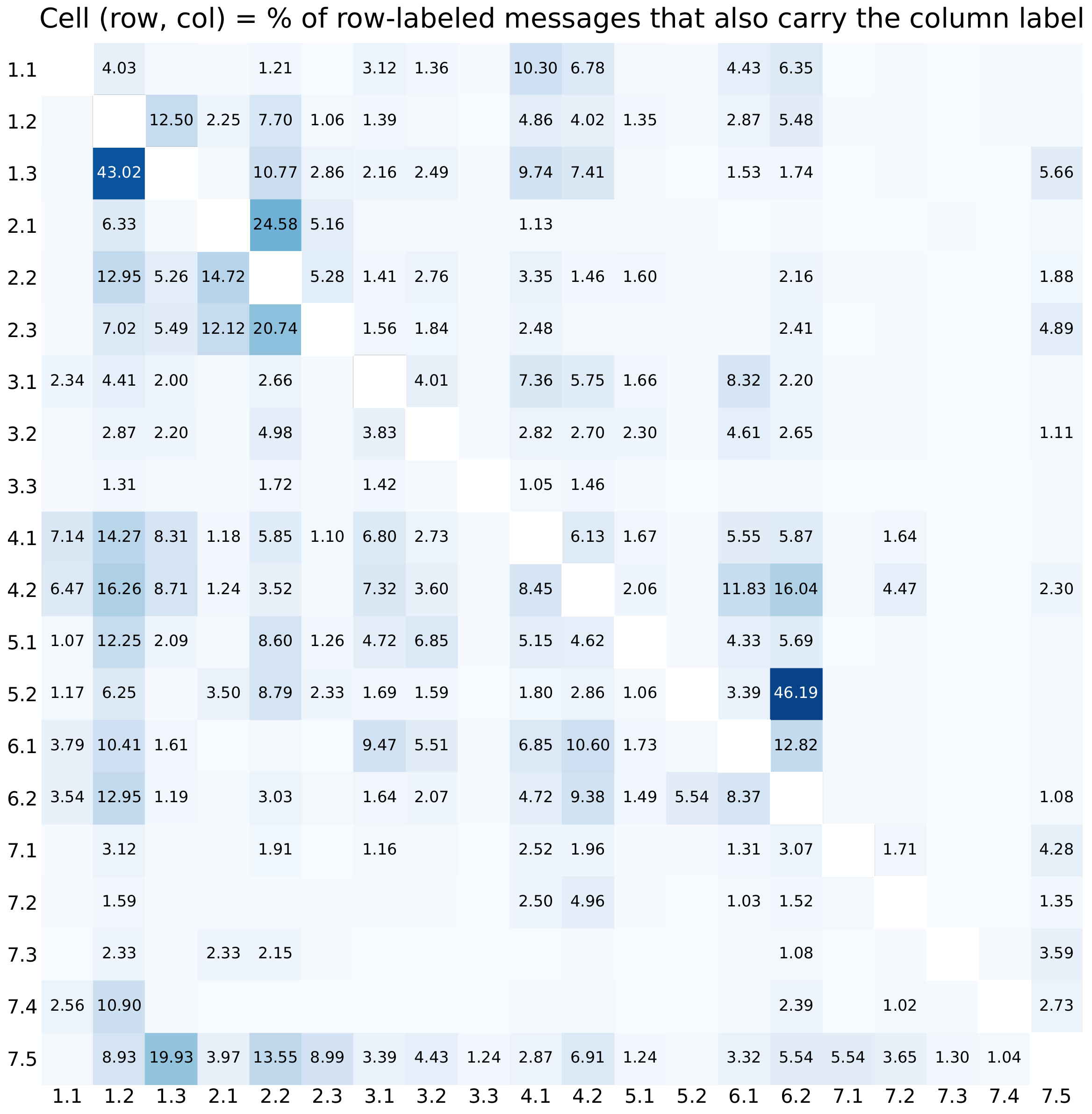}
\caption{Co-occurrence patterns among behavioral intent subcategories. Cell values show the percentage of messages carrying the row label that also carry the column label; values below 1\% are hidden. Diagonal cells are masked. Subcategory indices correspond to Table~\ref{tab:taxonomy}.}
\label{fig:cooccurrence}
\end{figure}

Table~\ref{tab:taxonomy} shows the overall distribution of behavioral intents. \textsc{Code Authoring} was the most prevalent category (34.53\%), followed by \textsc{Failure Reporting} (24.00\%), \textsc{Inquiry} (19.17\%), \textsc{Delegation} (16.48\%), \textsc{Context Specification} (14.08\%), and \textsc{Workflow Control} (11.47\%), whereas \textsc{Validation} was comparatively rare (3.99\%). Multi-intent prompting was common: 29.87\% of messages carried more than one label (mean = 1.33), though cases with three or more were rare (2.45\%). As shown in Figure~\ref{fig:cooccurrence}, co-occurring labels were structurally patterned rather than arbitrary. Multi-labeling rates were highest for \textsc{Alignment Correction} (83.63\%), \textsc{Behavior Specification} (77.88\%), and lowest for brief conversational control signals such as \textsc{Confirmation} (21.75\%) and \textsc{Continuation} (16.67\%). \added{Intent distributions were broadly consistent across the two largest language subsets, TypeScript/JavaScript/HTML (n = 40,472 messages) and Python (n = 13,537): all category differences fell within 8.5 percentage points, the largest being lower \textsc{Code Authoring} (27.82\% vs. 36.34\%) and higher \textsc{Delegation} (21.78\% vs. 15.28\%) in Python.}

\textbf{Finding 1. Conversational programming operates through progressive refinement rather than upfront specification: iterative modification and correction dominate over new implementation requests.}
Although \textsc{Code Authoring} was the most prevalent (34.53\%), \textsc{New Implementation} was rare (5.86\%), compared to \textsc{Iterative Modification} (24.84\%) and \textsc{Alignment Correction} (7.21\%), the latter co-occurring with \textsc{Iterative Modification} in 43.02\% of cases. Rather than issuing complete task descriptions upfront, developers predominantly revised and refined partially realized outputs, relying on accumulated context to keep correction turns minimal (e.g., \textit{``sans tiret''}, \textit{``Nope, still gray.''}).

This iterative pattern arose from two sources: realigning the assistant when its output diverged from intent (e.g., \textit{``Why aren't you following @user-preferences.mdc right now?''}), and accommodating evolving requirements as goals shifted (e.g., \textit{``Can we convert back to cmdliner?''}) or surrounding system changed (e.g., \textit{``I added a \texttt{course\_name} field on the backend; update the frontend accordingly''}).

\textbf{Finding 2. Developers delegated error diagnosis to AI by reporting symptoms and machine outputs rather than articulating code-level causes.}
\textsc{Failure Reporting} appeared in 24.00\% of all messages, making it the second most common category, with \textsc{Symptom Description} (14.77\%) more prevalent than \textsc{Log Paste} (8.84\%). When problems arose, developers commonly supplied raw machine outputs (compiler errors, runtime traces, terminal prints, etc.), described observable mismatches between expected and actual behavior (e.g., \textit{``table doesn't update until I refresh''}), or issued terse follow-ups (e.g., \textit{``still stuck''}) that depended almost entirely on prior context. Troubleshooting thus reflected a redistribution of diagnostic work: the assistant took on causal interpretation while developers acted primarily as reporters of failure evidence.

\textbf{Finding 3. When seeking to understand existing systems, developers queried AI about behavior, outputs, and feature logic, \added{with less emphasis on code semantics}.}
\textsc{Project Comprehension} accounted for 8.19\% of all messages. Prompts were commonly posed at the level of behavior or interpretation (e.g., \textit{``How does the system determine the icon's latitude and longitude?''}, \textit{``I don't understand the results---does this test now invalidate the hypothesis?''}), or used to verify developers' own interpretation of the system (e.g., \textit{``If I set \texttt{progress<=0.05} here, would that imply that \ldots?''}). Code-centric queries about what specific code does or how it is structured were comparatively rare (e.g., \textit{``what is calling \texttt{combinedUrlActions}?''}).

\textbf{Finding 4. Developers used AI to evaluate code through both static review and dynamic runtime inspection.} \textsc{Validation} appeared in 3.99\% of all messages, with \textsc{Code Review} (2.74\%) and \textsc{Runtime Inspection} (1.26\%) as its two components. Code review requests ranged from focused inspection to broader correctness checks (e.g., \textit{``Is the current implementation correct in this form?''}, \textit{``Make sure the modal is rendered properly on mobile''}), and frequently co-occurred with \textsc{Iterative Modification} (12.25\%) and \textsc{Symptom Description} (8.60\%), indicating that review was embedded in broader cycles of refinement and debugging.
\textsc{Runtime Inspection} extended this evaluative role to execution results (e.g., \textit{``Please run that command in a terminal so you can check the results''}), with nearly half of such messages (46.19\%) also carrying \textsc{Toolchain Operation}, suggesting that developers relied on the assistant not only to run code but also to evaluate its runtime behavior.

\textbf{Finding 5. Developers used AI-generated documentation to externalize plans, explanations, and progress into persistent artifacts for both future AI use and human understanding.}
\textsc{Documentation} appeared in 6.85\% of messages, commonly taking the form of Markdown-based planning, requirements, and progress files (e.g., \texttt{TODO.md}, \texttt{PROGRESS.md}). Developers created such artifacts primarily for future AI use: to guide implementation (e.g., \textit{``read \texttt{architecture.md} for context then fully implement''}), to validate code against written constraints (e.g., \textit{``check whether the implementation is consistent with the requirements and design documents''}), or to carry work across sessions (e.g., \textit{``create a prompt for the next chat to carry out all the next steps''}). They also produced human-facing artifacts such as module summaries, deployment audits, and architecture diagrams. Across both uses, documentation served as persistent external memory, stabilizing context and decisions across turns and sessions.

\textbf{Finding 6. Developers actively managed AI autonomy by shaping its context and constraining or delegating its behavior.}
\textsc{Information Injection} accounted for 8.46\% of all messages, providing context developers believed the assistant should incorporate, including relevant resources (e.g., \textit{``see the attached \texttt{openapi.yaml}''}), environment facts (e.g., \textit{``I've restarted this instance of VS Code since our last interaction so the env variable activated''}), and project-state updates (e.g., \textit{``I added two new states, \texttt{step\_dirty} and \texttt{code\_dirty}, to \texttt{index.d.ts}''}). Developers also explicitly set AI behavioral boundaries through \textsc{Behavior Specification} (6.14\%), including hard constraints (e.g., \textit{``only analyze; do not modify the code''}), conditions for human intervention (e.g., \textit{``if you encounter an error, stop and ask me''}), and, in some cases, open-ended delegation (e.g., \textit{``you decide''}), which traded direct oversight for reduced effort.

\subsection{RQ2. Session Archetypes and Dynamics}

We first characterize session-length distributions and recurring archetypes, then examine how behavioral intents evolve within and across sessions. Unless otherwise noted, group comparisons use Welch's $t$-test~\cite{welch1947generalization}, and all reported contrasts and regression coefficients were statistically significant ($p < 0.001$).

\subsubsection{Session Archetypes}
\label{sec:result_archetypes}

\begin{figure}[t]
\centering
\includegraphics[width=0.8\columnwidth]{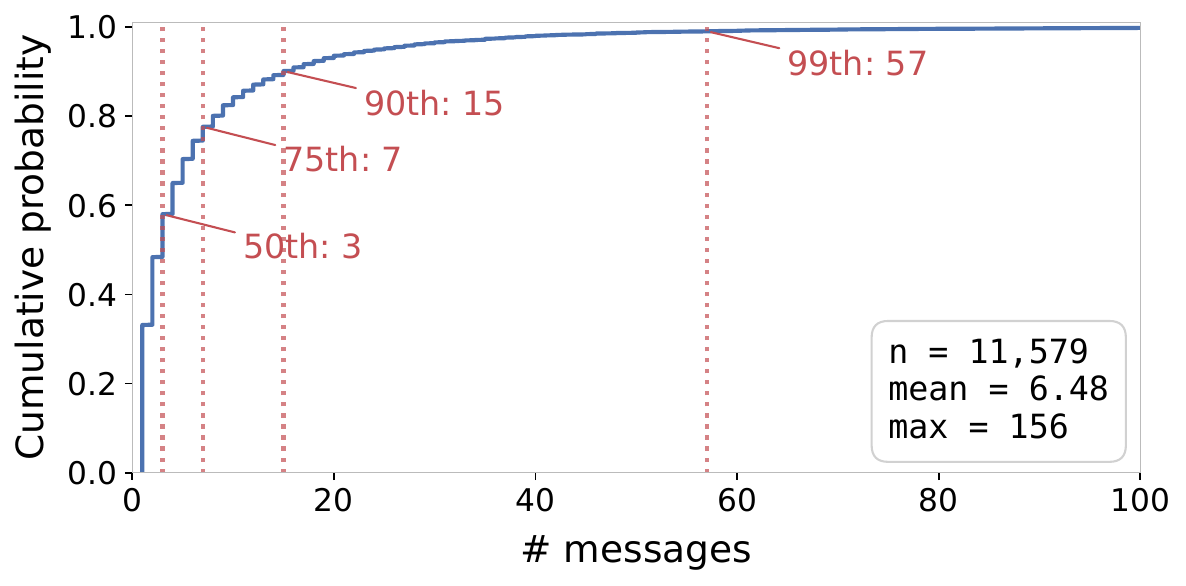}
\caption{Cumulative distribution of user messages per session ($n = 11{,}579$). Dotted lines mark selected percentiles.}
\label{fig:session_length}
\end{figure}

\textbf{Finding 7. Although most sessions were short and task-focused, a long tail of extended sessions centered on iterative refinement and debugging.}
The distribution of messages per session was right-skewed (median = 3; mean = 6.48; max = 156; Figure~\ref{fig:session_length}). Most sessions were short ($\leq$3 messages; $n = 6{,}715$) and concentrated in one-shot task-completion intents, such as \textsc{New Implementation} (12.78\% vs.\ 4.71\%), \textsc{Documentation} (9.75\% vs.\ 6.37\%), and \textsc{Deferred Debugging} (1.98\% vs.\ 0.54\%). Longer sessions ($\geq$4 messages; $n = 4{,}864$) instead showed higher proportions of reactive and iterative intents, including \textsc{Alignment Correction} (7.85\% vs.\ 3.41\%), \textsc{Error Persistence} (4.21\% vs.\ 1.06\%), and \textsc{Confirmation} (2.95\% vs.\ 0.86\%). We therefore restricted clustering to these longer sessions to ensure sufficient sequential signal for meaningful comparison (Section~\ref{sec:clustering}).

\begin{figure}[t]
\centering
\includegraphics[width=\columnwidth]{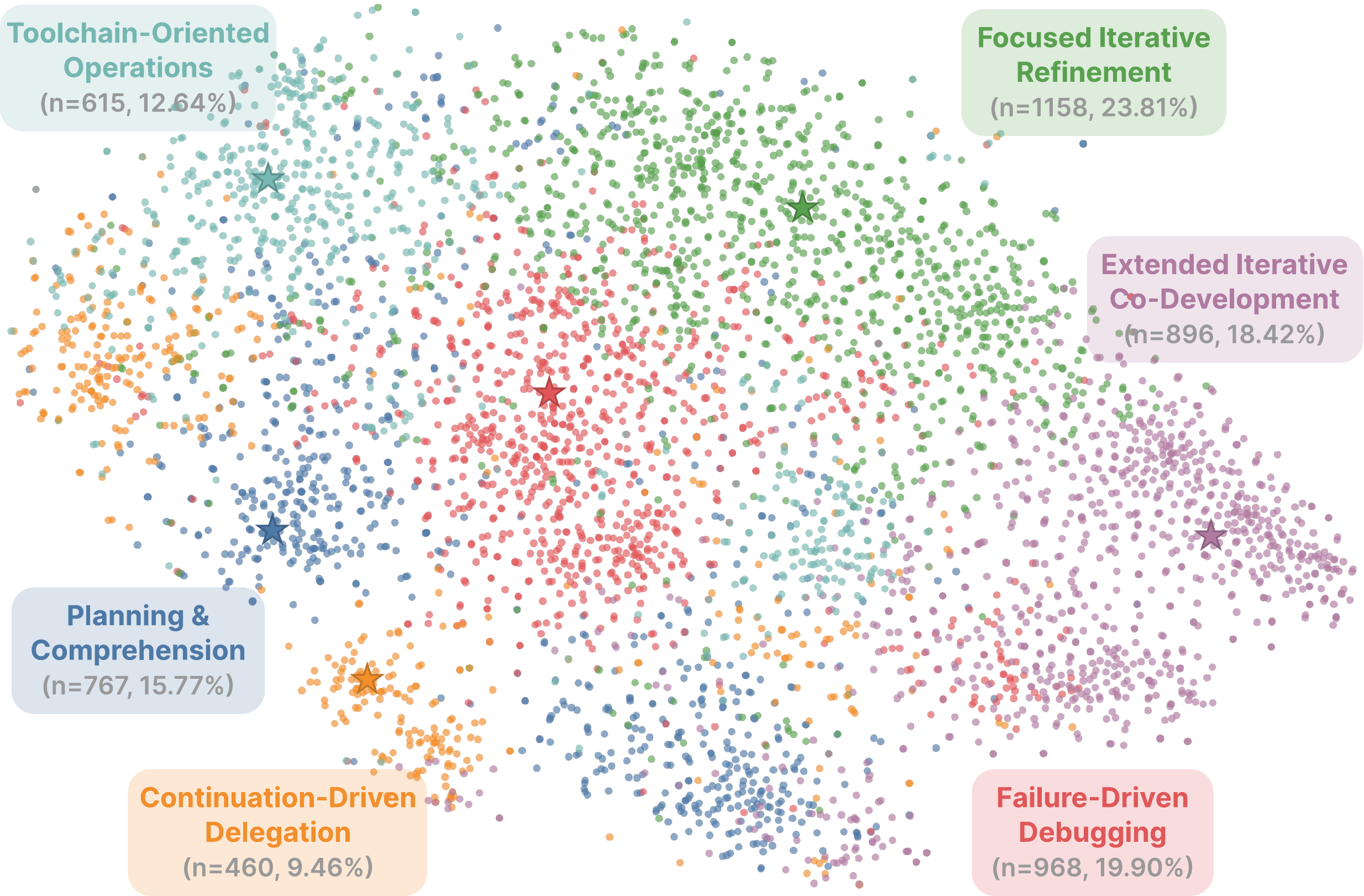}
\caption{Session archetype structure visualized as a t-SNE projection~\cite{van2008visualizing} of 4,864 sessions. Stars denote medoids.}
\label{fig:tsne}
\end{figure}

\begin{figure}[t]
\centering
\includegraphics[width=\columnwidth]{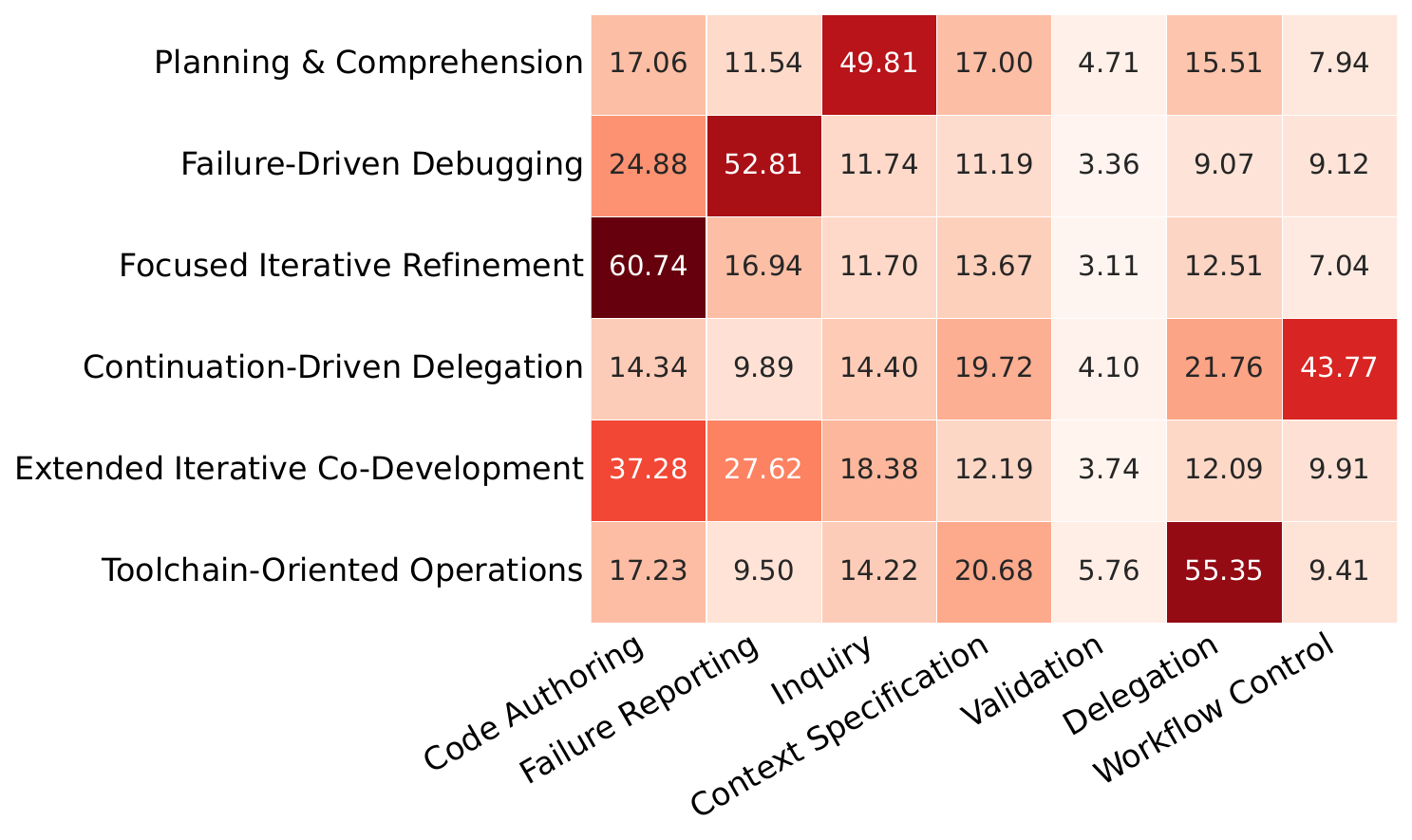}
\caption{Behavioral intent composition of each session archetype (\% of messages per main category).}
\label{fig:archetype_heatmap}
\end{figure}

\textbf{Finding 8. Clustering over longer sessions revealed six recurring session archetypes, each organized around a distinct behavioral mode.}
Figures~\ref{fig:tsne} and~\ref{fig:archetype_heatmap} show the t-SNE projection~\cite{van2008visualizing} and main-category distributions. Five archetypes consisted of short sessions (median: 6--8 messages), each organized around a single dominant category, whereas \textit{Extended Iterative Co-Development} stood apart with substantially longer sessions (median: 27) and the most balanced intent profile. We characterize each archetype below as descriptive prototypes of recurring interaction patterns rather than sharply separated clusters.

\paragraph{Planning \& Comprehension} ($n = 767$, 15.77\%).
Sessions were dominated by \textsc{Inquiry} (49.81\%), nearly evenly split between \textsc{Planning \& Consultation} (20.99\%) and \textsc{Project Comprehension} (20.37\%). Developers used the assistant as a thinking partner, asking how existing systems worked and exploring architectural options. The category distribution was stable across session positions, indicating that planning and comprehension functioned as a sustained interaction mode rather than a preliminary phase before coding.

\paragraph{Failure-Driven Debugging} ($n = 968$, 19.90\%).
Sessions were dominated by \textsc{Failure Reporting} (52.81\%), with \textsc{Symptom Description} (29.74\%) and \textsc{Log Paste} (25.00\%) as the primary subcategories. This archetype showed the clearest temporal progression among the six: sessions typically began with an \textsc{Iterative Modification} that triggered unexpected behavior, after which \textsc{Code Authoring} declined from 32.31\% to 20.03\% while \textsc{Failure Reporting} rose from 43.77\% to 59.86\%, indicating that an initial modification attempt gave way to sustained error-resolution cycles.

\paragraph{Focused Iterative Refinement} ($n = 1{,}158$, 23.81\%).
The largest archetype, with \textsc{Code Authoring} accounting for 60.74\% of messages, was driven primarily by \textsc{Iterative Modification} (48.98\%). Unlike \textit{Failure-Driven Debugging}, \textsc{Failure Reporting} appeared in only 16.94\% of messages, and the \textsc{Code Authoring} distribution remained stable across session positions (59.80\% to 63.14\%).

\paragraph{Continuation-Driven Delegation} ($n = 460$, 9.46\%).
\textsc{Workflow Control} was the dominant category (43.77\%), with \textsc{Continuation} alone accounting for 31.00\% of messages, the highest share of any subcategory across all archetypes. Sessions typically began with a substantive directive, often combining \textsc{Toolchain Operation} (13.32\%) and \textsc{Behavior Specification} (11.69\%), followed by a long sequence of continuation signals (e.g., \textit{``continue''}, \textit{``go on''}), suggesting that developers established the task and constraints upfront, then allowed the assistant to proceed with minimal intervention. \added{This highly delegative pattern resembles the ``vibe coding'' practice.}

\paragraph{Extended Iterative Co-Development} ($n = 896$, 18.42\%).
Distinguished primarily by session length (median: 27 messages; IQR: 20--40, compared with 6--8 in all other archetypes), this archetype also exhibited the most balanced intent profile, with no main category exceeding 37.3\% and no strong temporal trends, suggesting that these long sessions sustained a mixed-mode collaborative process across a full development cycle. \added{Qualitative inspection revealed recurring patterns, e.g., developers interleaved implementation, debugging, and validation within a single session, often across multiple work contexts.}

\paragraph{Toolchain-Oriented Operations} ($n = 615$, 12.64\%).
Sessions were centered on \textsc{Delegation} (55.35\%), driven by \textsc{Toolchain Operation} (38.55\%) and \textsc{Documentation} (20.50\%), with \textsc{Context Specification} elevated (20.68\%) to supply environment details (e.g., port numbers, file paths). Developers instruct the assistant to perform environment and project operations, e.g., starting servers, managing Git branches, installing dependencies, and generating documentation, rather than to produce or debug application code.

\subsubsection{Intent Dynamics}
\label{sec:result_dynamics}

\begin{figure}[t]
\centering
\includegraphics[width=\columnwidth]{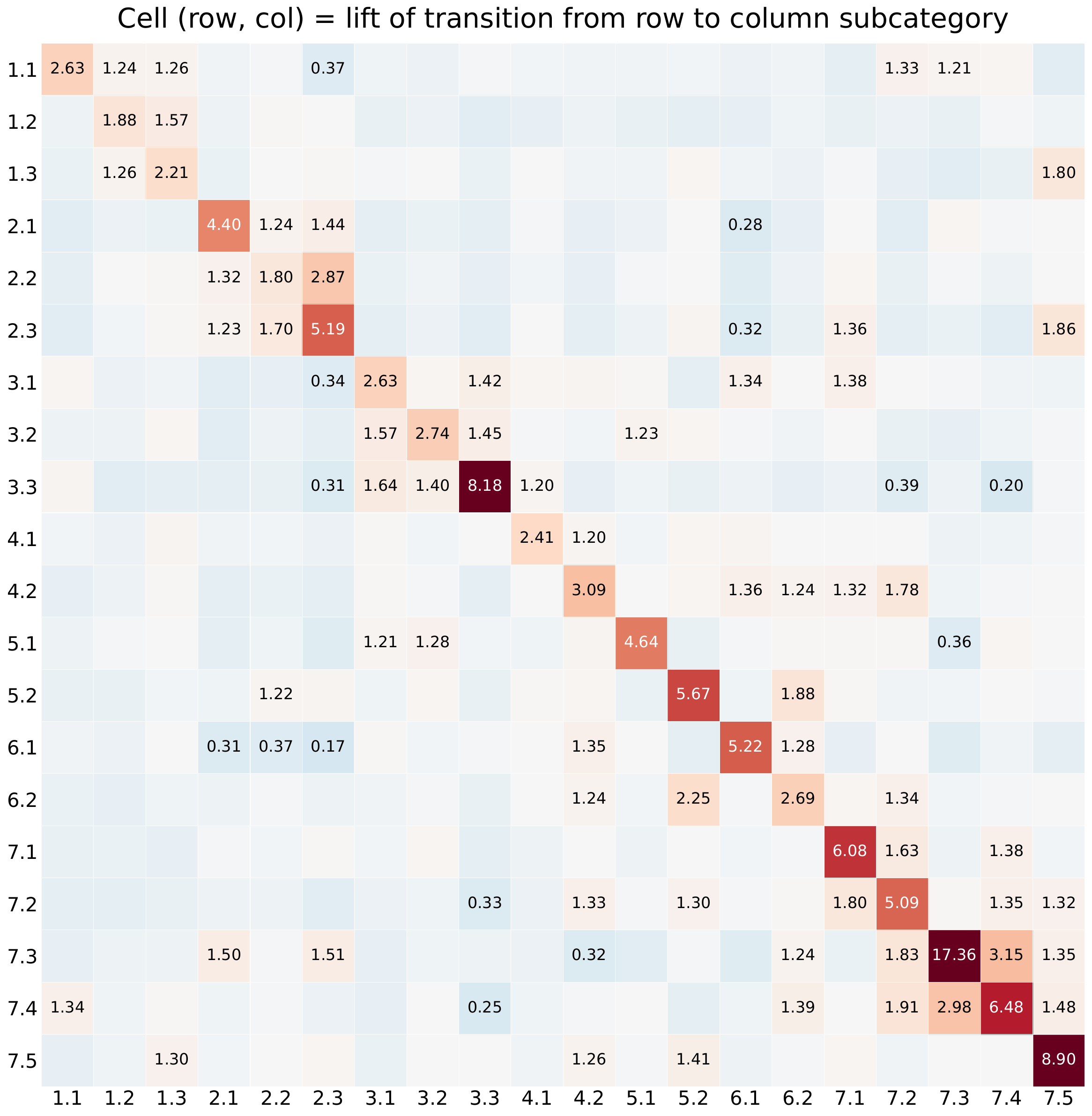}
\caption{Transition patterns between behavioral intent subcategories within sessions, measured as Markov lift. Diagonal entries represent self-loop lift. Numerical values are shown for notably reinforced (lift $\geq 1.2$) or suppressed (lift $\leq 0.4$) transitions; all other cells display color only. Subcategory indices correspond to Table~\ref{tab:taxonomy}.}
\label{fig:transition_lift}
\end{figure}

\textbf{Finding 9. Within-session transitions were strongly self-reinforcing, producing sustained runs of iterative modification and failure reporting, and recurrent micro-cycles centered on debugging and runtime validation.}
All 20 subcategories exhibited self-loop lift (min = 1.80, max = 17.36\added{; Figure~\ref{fig:transition_lift}}). Measured as consecutive turns with the same intent, \textsc{Iterative Modification} showed the strongest continuity (mean run length = 1.57; 29.96\% multi-turn runs; longest run = 27 turns), followed by \textsc{Log Paste} (1.43; 25.01\%; 13 turns).
Beyond self-loops, several cross-category transitions formed recurrent micro-cycles: a debugging loop linked \textsc{Symptom Description} $\rightarrow$ \textsc{Error Persistence} (lift = 2.87) and back (lift = 1.70); a validation loop connected \textsc{Toolchain Operation} and \textsc{Runtime Inspection} bidirectionally (lift = 2.25 and 1.88); and \textsc{Iterative Modification} frequently transitioned to \textsc{Alignment Correction} (lift = 1.57), reflecting corrective redirection when the assistant's output diverged from intent.

\textbf{Finding 10. Session boundaries preserved task-level intent continuity while resetting conversational states, suggesting that developers used session breaks to refresh accumulated context rather than to switch tasks.}
Across session boundaries, self-loop lifts remained above 1 in all subcategories (min = 1.19, max = 20.77), indicating preserved task-level intents. However, context-dependent subcategories exhibited substantial declines: \textsc{Confirmation} dropped from a within-session lift of 6.08 to nearly zero, and \textsc{Error Persistence} declined from 5.19 to 1.19.

\textbf{Finding 11. Sessions are not structurally uniform: developers begin by establishing task scopes and constraints, then progressively shift toward responding to intermediate AI outputs and emerging failures.}
Opening messages served a distinct framing function. They were roughly twice as long as later turns (1,003 vs.\ 499 characters) and disproportionately featured setup-oriented acts, including \textsc{New Implementation} (15.89\% vs.\ 4.03\%), \textsc{Information Injection} (11.82\% vs.\ 7.84\%), and \textsc{Behavior Specification} (7.46\% vs.\ 5.90\%). In contrast, intents that depend on prior conversational context, such as \textsc{Alignment Correction} (0.73\% vs.\ 8.40\%) and \textsc{Continuation} (1.73\% vs.\ 6.24\%), were nearly absent.\looseness=-1

Beyond the opening turn, sessions became increasingly reactive to AI outputs. \textsc{Inquiry} and \textsc{Context Specification} declined monotonically across session position ($R^2 = 0.70$ and $0.90$), while \textsc{Failure Reporting} and \textsc{Delegation} increased ($R^2 = 0.66$ and $0.61$). Message length also grew substantially (slope $= 198.4$, $R^2 = 0.91$), whereas labels per message remained nearly flat ($R^2 = 0.02$). Notably, \textsc{Continuation}, nearly absent at session onset, peaked in the early post-opening phase before declining steadily ($R^2 = 0.91$), suggesting that developers first established the task, then briefly sustained momentum before shifting to reactive engagement.

\section{Threats to Validity}

\textit{External validity.} Our dataset reflects developers who use SpecStory and commit chat histories to public GitHub repositories, introducing selection bias \added{that may underrepresent private and enterprise-level projects}. Rather than approximating the full developer population, this dataset more directly reflects early adopters and heavy users of conversational programming workflows. The dataset is further scoped to IDE-integrated assistants and may not generalize to other forms of AI-assisted programming (e.g., CLI-based agents). At the same time, the naturalistic and unsolicited nature of the data avoids the observation bias common in lab or task-assigned settings. \added{Notably, survivorship bias, i.e., developers preferentially sharing successful sessions, is partially mitigated by SpecStory's design: histories are auto-exported in batches with little manual curation, and our data contains substantial unresolved-failure signals that strong survivorship filtering would underrepresent.} As model capabilities and user habits evolve rapidly, our results should be interpreted as a snapshot of early-stage conversational programming practice rather than stable long-term distributions.

\textit{Data completeness and construct validity.} Our analysis is limited to text-based developer behavior preserved in SpecStory exports. Non-textual attachments (e.g., images or binary files), inline code references, and UI-level actions (e.g., regeneration, cancellation, code accept/reject) are not consistently recorded. Consequently, non-text user behaviors expressed through the interface, particularly \textsc{Information Injection}, are not represented. More broadly, our taxonomy captures developers' expressed behavioral intents, not their downstream code quality, correctness, or productivity effects.\looseness=-1

\textit{Internal validity.} Several stages of the analysis involve methodological choices that may influence the results. Codebook development and qualitative interpretation involve subjective judgment of researchers, which we mitigated through multi-researcher coding, iterative refinement, and consensus discussion. The LLM classifier was validated on a stratified sample covering all 20 subcategories. While some classification errors may remain and could affect downstream frequency estimates, our main findings rely on aggregate distributions and recurring patterns unlikely to be driven by isolated misclassifications.
Other design choices, including sampling sizes, context truncation, minimum session length for clustering, and edit-distance parameterization, are theoretically motivated~\cite{abbott2000sequence,studer2016matters} but non-unique; alternative specifications may yield different distributions or archetypes. We therefore interpret the clusters as descriptive archetypes rather than sharply bounded classes.
\section{Discussion}

\subsection{Conversational Programming as Progressive Specification}

Classical requirements engineering treats specification as a precondition for implementation. Conversational programming disrupts this sequencing: \textsc{Alignment Correction} and \textsc{Symptom Description} not only fix broken code but also introduce previously unarticulated constraints, distributing requirements articulation across the dialogue rather than concentrating it upfront (Finding 1, 11).

This underspecified communication is consistent with the bimodal distribution of message length: 33.77\% of messages were 50 characters or shorter, while 9.02\% exceeded 1,000 characters. At one extreme, terse cues such as \textit{``Nope, still gray.''} carried meaning only through context; at the other, developers embedded raw logs (\textsc{Log Paste}, median: 718 characters) or lengthy context blocks (\textsc{Information Injection}, median: 162 characters), leaving the assistant to determine what was relevant. This pattern likely differs from browser-based chatbot interactions~\cite{xiao2024devgpt, zhong2025developer}: whereas both accumulate conversational history, IDE-native assistants additionally have direct access to the full codebase, runtime outputs, and development environment, and are expected to actively gather relevant context through file reads, codebase search, and command execution rather than relying on what the developer has made explicit.\looseness=-1

One response to this challenge is to externalize plans into persistent documents, a practice formalized as \textit{spec-driven development}: tools such as GitHub Spec Kit\footnote{\url{https://github.github.com/spec-kit/}} and Amazon Kiro\footnote{\url{https://aws.amazon.com/documentation-overview/kiro/}} treat lightweight documents (e.g., \texttt{design.md}, \texttt{tasks.md}) as the source of truth guiding AI agents through implementation. Our data suggest developers are already doing this organically: \textsc{Documentation} appeared in 6.85\% of messages, with developers creating planning documents to record intent and progress across turns and sessions (Finding 5). However, because these documents are themselves generated from underspecified prompts, they introduce a potential compounded validation problem: developers must assess not only whether the code is correct, but whether the document captures their intent and whether the assistant's actions are consistent with it, a three-way alignment problem that warrants further investigation.

These findings carry direct implications for benchmark design. Dominant benchmarks such as SWE-bench~\cite{jimenez2023swe} and HumanEval~\cite{chen2021evaluating} assume a complete, self-contained task description, an assumption systematically violated in IDE-native use where specifications emerge over multiple turns from underspecified signals. A model optimized under this assumption may perform well on benchmarks while failing at tolerating ambiguity, tracking evolving intent, and proactively surfacing missing constraints. The emergence of CursorBench\footnote{\url{https://cursor.com/blog/cursorbench}}, which explicitly emphasizes underspecified tasks, suggests that tool builders are beginning to recognize this gap.

\subsection{Distributed Cognition in Debugging, Comprehension, and Validation}

Conversational programming appears to redistribute cognitive labor from developer to assistant: tasks such as fault diagnosis, code comprehension, and output validation are increasingly offloaded to AI, while developers retain the roles of symptom reporter, intent clarifier, and final evaluator. This redistribution is most visible in debugging. Although \textsc{Failure Reporting} was the second most prevalent category, developers rarely articulated code-level causes (Finding~2). Instead, they often relied on terse signals such as \textit{``still not working''}, leaving diagnosis almost entirely to the assistant. These minimal messages function as low-cost probes: if the assistant recovers, the developer avoids the effort of diagnosis; if not, additional detail can be provided incrementally. This norm differs markedly from bug reporting to human collaborators, where detailed context signals both competence and respect for others' time. Sentiment data reinforced this interpretation: negative affect co-occurred most frequently with \textsc{Alignment Correction} (24.20\%), \textsc{Symptom Description} (15.58\%), and \textsc{Error Persistence} (11.01\%), suggesting that emotional expressions serve less as social commentary than as compressed signals of unresolved failure.

As AI generates code faster than developers can inspect it, comprehension and validation are also displaced onto the assistant. Rather than reading implementations directly, developers queried the assistant about system behavior and feature logic to reconstruct how the code worked from the outside in (Finding~3). Validation followed the same pattern: developers delegated both static code review and runtime inspection to the assistant (Finding~4). This shift reduces the immediate burden of engaging with large volumes of generated code, but raises a deeper concern: when diagnosis, comprehension, and validation are all delegated to the same system, the assistant may become the only judge of both what the code does and whether it is correct, with no independent check on its own errors.\looseness=-1

\subsection{Managing an Opaque Collaborator}

Conversational programming introduces a new coordination problem: although the state of the collaboration is nominally preserved in the chat history, it is difficult for developers to access in practice. In agentic IDE sessions, file reads, terminal commands, code edits, and long textual responses accumulate in a single stream, making it hard to monitor what the assistant knows and what it has done.

This opacity creates two related problems. First, developers may struggle to validate what the assistant has already done. To compensate, they issued explicit status checks (e.g., \textit{``is your task finished?''}) and asked the assistant to externalize its own actions (e.g., \textit{``Write down the changes in \texttt{PROGRESS.md} for me to track''}) (Finding~5). In effect, developers relied on assistant self-report because the interaction history had become too long and dense to inspect directly.

Second, developers may not reliably steer what the assistant should do next. To manage this, they injected missing context and imposed behavioral constraints, ranging from explicit hard stops to open-ended handoffs (Finding~6). The \textit{Continuation-Driven Delegation} archetype represents the strongest form of the latter. As agentic systems take on longer task horizons, this coordination burden is likely to increase.

\subsection{Future Work}

Several directions remain beyond the scope of this study. First, our analysis captures what developers do but not whether those interactions achieved their goals; addressing this would require retrospective accounts of which requests and responses developers regarded as failures and why. Second, we exclude CLI-based agent sessions (e.g., Claude Code), which operate with greater autonomy and less turn-by-turn human direction; comparing conversational and more agentic workflows could clarify how increasing AI autonomy affects developer-AI interaction. Third, conversational programming may demand skills underemphasized in traditional programming education~\cite{yang2025spark, yang2026lessons}, including requirement articulation, AI output evaluation, and context management, warranting curricular investigation. \added{Finally, breakdowns by organization, domain, and time period are limited by our predominantly single-contributor, hobby-skewed sample and the tools' recent emergence. Extending such comparisons using our public taxonomy is a promising direction.}
\section{Conclusion}

We presented the first large-scale empirical study of AI-assisted conversational programming in IDE-native environments, characterizing developers' behavioral 
intents and recurring session-level patterns from real-world developer-AI conversations. We hope this empirical foundation informs future AI coding assistants that better support iterative, context-dependent developer-AI collaboration.



\section*{Data Availability Statement}

The replication package, including the codebook, classification prompts, and all analysis code, is available on GitHub\footnote{\url{https://github.com/ND-SaNDwichLAB/empirical-conversational-programming}} and archived at \href{https://doi.org/10.5281/zenodo.19248043}{\raisebox{-0.25em}{\includegraphics[scale=0.7]{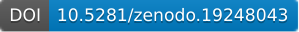}}}. Raw chat-session data are excluded due to copyright and privacy considerations, as most source repositories do not carry explicit licenses permitting redistribution. However, the data can be fully recollected via the provided scraping pipeline\footnote{\url{https://github.com/TTangNingzhi/vibe-coding-scraper}}. Researchers interested in obtaining direct access may contact the first author.

\begin{acks}
This research was supported in part by a Google Cloud Research Credit Award, a Google Research Scholar Award, a gift from Adobe, an Edison Innovation Fellowship from the Notre Dame IDEA Center, and NSF grants CCF-2211428, CCF-2315887, CCF-2100035, CCF-2211429, CCF-2442682, and DGE-2622415. Any opinions, findings, or recommendations expressed here are those of the authors and do not necessarily reflect the views of the sponsors. The authors thank Yuqi Wang from CREVIK for introducing us to SpecStory, without which this study would not have been possible.
\end{acks}

\balance
\bibliographystyle{ACM-Reference-Format}
\bibliography{refs}


\begin{thebibliography}{67}


\ifx \showCODEN    \undefined \def \showCODEN     #1{\unskip}     \fi
\ifx \showISBNx    \undefined \def \showISBNx     #1{\unskip}     \fi
\ifx \showISBNxiii \undefined \def \showISBNxiii  #1{\unskip}     \fi
\ifx \showISSN     \undefined \def \showISSN      #1{\unskip}     \fi
\ifx \showLCCN     \undefined \def \showLCCN      #1{\unskip}     \fi
\ifx \shownote     \undefined \def \shownote      #1{#1}          \fi
\ifx \showarticletitle \undefined \def \showarticletitle #1{#1}   \fi
\ifx \showURL      \undefined \def \showURL       {\relax}        \fi
\providecommand\bibfield[2]{#2}
\providecommand\bibinfo[2]{#2}
\providecommand\natexlab[1]{#1}
\providecommand\showeprint[2][]{arXiv:#2}

\bibitem[Abbott and Tsay(2000)]%
        {abbott2000sequence}
\bibfield{author}{\bibinfo{person}{Andrew Abbott} {and} \bibinfo{person}{Angela Tsay}.} \bibinfo{year}{2000}\natexlab{}.
\newblock \showarticletitle{Sequence analysis and optimal matching methods in sociology: Review and prospect}.
\newblock \bibinfo{journal}{\emph{Sociological methods \& research}} \bibinfo{volume}{29}, \bibinfo{number}{1} (\bibinfo{year}{2000}), \bibinfo{pages}{3--33}.
\newblock
\href{https://doi.org/10.1177/0049124100029001001}{doi:\nolinkurl{10.1177/0049124100029001001}}


\bibitem[Alaboudi and LaToza(2019)]%
        {alaboudi2019exploratory}
\bibfield{author}{\bibinfo{person}{Abdulaziz Alaboudi} {and} \bibinfo{person}{Thomas~D LaToza}.} \bibinfo{year}{2019}\natexlab{}.
\newblock \showarticletitle{An exploratory study of live-streamed programming}. In \bibinfo{booktitle}{\emph{2019 IEEE Symposium on Visual Languages and Human-Centric Computing (VL/HCC)}}. IEEE, \bibinfo{pages}{5--13}.
\newblock
\href{https://doi.org/10.1109/VLHCC.2019.8818832}{doi:\nolinkurl{10.1109/VLHCC.2019.8818832}}


\bibitem[AlOmar et~al\mbox{.}(2024)]%
        {alomar2024refactor}
\bibfield{author}{\bibinfo{person}{Eman~Abdullah AlOmar}, \bibinfo{person}{Anushkrishna Venkatakrishnan}, \bibinfo{person}{Mohamed~Wiem Mkaouer}, \bibinfo{person}{Christian Newman}, {and} \bibinfo{person}{Ali Ouni}.} \bibinfo{year}{2024}\natexlab{}.
\newblock \showarticletitle{How to refactor this code? an exploratory study on developer-chatgpt refactoring conversations}. In \bibinfo{booktitle}{\emph{Proceedings of the 21st International Conference on Mining Software Repositories}}. \bibinfo{pages}{202--206}.
\newblock
\href{https://doi.org/10.1145/3643991.3645081}{doi:\nolinkurl{10.1145/3643991.3645081}}


\bibitem[Arthur et~al\mbox{.}(2007)]%
        {arthur2006k}
\bibfield{author}{\bibinfo{person}{David Arthur}, \bibinfo{person}{Sergei Vassilvitskii}, {et~al\mbox{.}}} \bibinfo{year}{2007}\natexlab{}.
\newblock \showarticletitle{k-means++: The advantages of careful seeding}. In \bibinfo{booktitle}{\emph{Soda}}, Vol.~\bibinfo{volume}{7}. \bibinfo{pages}{1027--1035}.
\newblock
\showISBNx{9780898716245}


\bibitem[Aubakirova et~al\mbox{.}(2026)]%
        {aubakirova2026state}
\bibfield{author}{\bibinfo{person}{Malika Aubakirova}, \bibinfo{person}{Alex Atallah}, \bibinfo{person}{Chris Clark}, \bibinfo{person}{Justin Summerville}, {and} \bibinfo{person}{Anjney Midha}.} \bibinfo{year}{2026}\natexlab{}.
\newblock \showarticletitle{State of AI: An Empirical 100 Trillion Token Study with OpenRouter}.
\newblock \bibinfo{journal}{\emph{arXiv preprint arXiv:2601.10088}} (\bibinfo{year}{2026}).
\newblock
\href{https://doi.org/10.48550/arXiv.2601.10088}{doi:\nolinkurl{10.48550/arXiv.2601.10088}}


\bibitem[Austin(1975)]%
        {austin1975things}
\bibfield{author}{\bibinfo{person}{John~Langshaw Austin}.} \bibinfo{year}{1975}\natexlab{}.
\newblock \bibinfo{booktitle}{\emph{How to do things with words}}.
\newblock \bibinfo{publisher}{Harvard university press}.
\newblock


\bibitem[Barke et~al\mbox{.}(2023)]%
        {barke2023grounded}
\bibfield{author}{\bibinfo{person}{Shraddha Barke}, \bibinfo{person}{Michael~B James}, {and} \bibinfo{person}{Nadia Polikarpova}.} \bibinfo{year}{2023}\natexlab{}.
\newblock \showarticletitle{Grounded copilot: How programmers interact with code-generating models}.
\newblock \bibinfo{journal}{\emph{Proceedings of the ACM on Programming Languages}} \bibinfo{volume}{7}, \bibinfo{number}{OOPSLA1} (\bibinfo{year}{2023}), \bibinfo{pages}{85--111}.
\newblock
\href{https://doi.org/10.1145/3586030}{doi:\nolinkurl{10.1145/3586030}}


\bibitem[Chatterji et~al\mbox{.}(2025)]%
        {chatterji2025people}
\bibfield{author}{\bibinfo{person}{Aaron Chatterji}, \bibinfo{person}{Thomas Cunningham}, \bibinfo{person}{David~J Deming}, \bibinfo{person}{Zoe Hitzig}, \bibinfo{person}{Christopher Ong}, \bibinfo{person}{Carl~Yan Shan}, {and} \bibinfo{person}{Kevin Wadman}.} \bibinfo{year}{2025}\natexlab{}.
\newblock \bibinfo{booktitle}{\emph{How people use chatgpt}}.
\newblock \bibinfo{type}{{T}echnical {R}eport}. \bibinfo{institution}{National Bureau of Economic Research}.
\newblock
\href{https://doi.org/10.3386/w34255}{doi:\nolinkurl{10.3386/w34255}}


\bibitem[Chen et~al\mbox{.}(2021)]%
        {chen2021evaluating}
\bibfield{author}{\bibinfo{person}{Mark Chen}, \bibinfo{person}{Jerry Tworek}, \bibinfo{person}{Heewoo Jun}, \bibinfo{person}{Qiming Yuan}, \bibinfo{person}{Henrique Ponde De~Oliveira Pinto}, \bibinfo{person}{Jared Kaplan}, \bibinfo{person}{Harri Edwards}, \bibinfo{person}{Yuri Burda}, \bibinfo{person}{Nicholas Joseph}, \bibinfo{person}{Greg Brockman}, {et~al\mbox{.}}} \bibinfo{year}{2021}\natexlab{}.
\newblock \showarticletitle{Evaluating large language models trained on code}.
\newblock \bibinfo{journal}{\emph{arXiv preprint arXiv:2107.03374}} (\bibinfo{year}{2021}).
\newblock
\href{https://doi.org/10.48550/arXiv.2107.03374}{doi:\nolinkurl{10.48550/arXiv.2107.03374}}


\bibitem[Chen et~al\mbox{.}(2026)]%
        {chen2025code}
\bibfield{author}{\bibinfo{person}{Valerie Chen}, \bibinfo{person}{Ameet Talwalkar}, \bibinfo{person}{Robert Brennan}, {and} \bibinfo{person}{Graham Neubig}.} \bibinfo{year}{2026}\natexlab{}.
\newblock \showarticletitle{Code with me or for me? how increasing ai automation transforms developer workflows}. In \bibinfo{booktitle}{\emph{Proceedings of the 2026 CHI Conference on Human Factors in Computing Systems}}. \bibinfo{pages}{1--19}.
\newblock
\href{https://doi.org/10.1145/3772318.3790850}{doi:\nolinkurl{10.1145/3772318.3790850}}


\bibitem[Chi et~al\mbox{.}(2026)]%
        {chi2025edit}
\bibfield{author}{\bibinfo{person}{Wayne Chi}, \bibinfo{person}{Valerie Chen}, \bibinfo{person}{Ryan Shar}, \bibinfo{person}{Aditya Mittal}, \bibinfo{person}{Jenny Liang}, \bibinfo{person}{Wei-Lin Chiang}, \bibinfo{person}{Anastasios Angelopoulos}, \bibinfo{person}{Ion Stoica}, \bibinfo{person}{Graham Neubig}, \bibinfo{person}{Ameet Talwalkar}, {et~al\mbox{.}}} \bibinfo{year}{2026}\natexlab{}.
\newblock \showarticletitle{EditBench: Evaluating LLM Abilities to Perform Real-World Instructed Code Edits}. In \bibinfo{booktitle}{\emph{International Conference on Learning Representations}}, Vol.~\bibinfo{volume}{2026}. \bibinfo{pages}{94809--94835}.
\newblock


\bibitem[Chou et~al\mbox{.}(2026)]%
        {chou2025building}
\bibfield{author}{\bibinfo{person}{Yi-Hung Chou}, \bibinfo{person}{Boyuan Jiang}, \bibinfo{person}{Yi~Wen Chen}, \bibinfo{person}{Mingyue Weng}, \bibinfo{person}{Victoria Jackson}, \bibinfo{person}{Thomas Zimmermann}, {and} \bibinfo{person}{James~A Jones}.} \bibinfo{year}{2026}\natexlab{}.
\newblock \showarticletitle{Building software by rolling the dice: A qualitative study of vibe coding}.
\newblock \bibinfo{journal}{\emph{Proceedings of the ACM on Software Engineering}} \bibinfo{volume}{3}, \bibinfo{number}{FSE} (\bibinfo{year}{2026}), \bibinfo{pages}{1712--1734}.
\newblock
\href{https://doi.org/10.1145/3797105}{doi:\nolinkurl{10.1145/3797105}}


\bibitem[Costa-Gomes et~al\mbox{.}(2025)]%
        {costa2025s}
\bibfield{author}{\bibinfo{person}{Beatriz Costa-Gomes}, \bibinfo{person}{Sophia Chen}, \bibinfo{person}{Connie Hsueh}, \bibinfo{person}{Deborah Morgan}, \bibinfo{person}{Philipp Schoenegger}, \bibinfo{person}{Yash Shah}, \bibinfo{person}{Sam Way}, \bibinfo{person}{Yuki Zhu}, \bibinfo{person}{Timoth{\'e} Adeline}, \bibinfo{person}{Michael Bhaskar}, {et~al\mbox{.}}} \bibinfo{year}{2025}\natexlab{}.
\newblock \showarticletitle{It's About Time: The Temporal and Modal Dynamics of Copilot Usage}.
\newblock \bibinfo{journal}{\emph{arXiv preprint arXiv:2512.11879}} (\bibinfo{year}{2025}).
\newblock
\href{https://doi.org/10.48550/arXiv.2512.11879}{doi:\nolinkurl{10.48550/arXiv.2512.11879}}


\bibitem[Cui et~al\mbox{.}(2026)]%
        {cui2026effects}
\bibfield{author}{\bibinfo{person}{Kevin~Zheyuan Cui}, \bibinfo{person}{Mert Demirer}, \bibinfo{person}{Sonia Jaffe}, \bibinfo{person}{Leon Musolff}, \bibinfo{person}{Sida Peng}, {and} \bibinfo{person}{Tobias Salz}.} \bibinfo{year}{2026}\natexlab{}.
\newblock \showarticletitle{The effects of generative AI on high-skilled work: Evidence from three field experiments with software developers}.
\newblock \bibinfo{journal}{\emph{Management Science}} (\bibinfo{year}{2026}).
\newblock
\href{https://doi.org/10.1287/mnsc.2025.00535}{doi:\nolinkurl{10.1287/mnsc.2025.00535}}


\bibitem[Deng et~al\mbox{.}(2025)]%
        {deng2025swe}
\bibfield{author}{\bibinfo{person}{Xiang Deng}, \bibinfo{person}{Jeff Da}, \bibinfo{person}{Edwin Pan}, \bibinfo{person}{Yannis~Yiming He}, \bibinfo{person}{Charles Ide}, \bibinfo{person}{Kanak Garg}, \bibinfo{person}{Niklas Lauffer}, \bibinfo{person}{Andrew Park}, \bibinfo{person}{Nitin Pasari}, \bibinfo{person}{Chetan Rane}, {et~al\mbox{.}}} \bibinfo{year}{2025}\natexlab{}.
\newblock \showarticletitle{Swe-bench pro: Can ai agents solve long-horizon software engineering tasks?}
\newblock \bibinfo{journal}{\emph{arXiv preprint arXiv:2509.16941}} (\bibinfo{year}{2025}).
\newblock
\href{https://doi.org/10.48550/arXiv.2509.16941}{doi:\nolinkurl{10.48550/arXiv.2509.16941}}


\bibitem[Fang et~al\mbox{.}(2025a)]%
        {fang2025comparative}
\bibfield{author}{\bibinfo{person}{Zihan Fang}, \bibinfo{person}{Jiliang Li}, \bibinfo{person}{Anda Liang}, \bibinfo{person}{Gina~R Bai}, {and} \bibinfo{person}{Yu Huang}.} \bibinfo{year}{2025}\natexlab{a}.
\newblock \showarticletitle{A comparative study on chatgpt and checklist as support tools for unit testing education}. In \bibinfo{booktitle}{\emph{Proceedings of the 33rd ACM International Conference on the Foundations of Software Engineering}}. \bibinfo{pages}{871--882}.
\newblock
\href{https://doi.org/10.1145/3696630.3727244}{doi:\nolinkurl{10.1145/3696630.3727244}}


\bibitem[Fang et~al\mbox{.}(2025b)]%
        {fang2025dpo}
\bibfield{author}{\bibinfo{person}{Zihan Fang}, \bibinfo{person}{Yifan Zhang}, \bibinfo{person}{Yueke Zhang}, \bibinfo{person}{Kevin Leach}, {and} \bibinfo{person}{Yu Huang}.} \bibinfo{year}{2025}\natexlab{b}.
\newblock \showarticletitle{DPO-F+: Aligning Code Repair Feedback with Developers' Preferences}.
\newblock \bibinfo{journal}{\emph{arXiv preprint arXiv:2511.01043}} (\bibinfo{year}{2025}).
\newblock
\href{https://doi.org/10.48550/arXiv.2511.01043}{doi:\nolinkurl{10.48550/arXiv.2511.01043}}


\bibitem[Fawzy et~al\mbox{.}(2026)]%
        {fawzy2025vibe}
\bibfield{author}{\bibinfo{person}{Ahmed Fawzy}, \bibinfo{person}{Amjed Tahir}, {and} \bibinfo{person}{Kelly Blincoe}.} \bibinfo{year}{2026}\natexlab{}.
\newblock \showarticletitle{Vibe coding in practice: Motivations, challenges, and a future outlook--a grey literature review}. In \bibinfo{booktitle}{\emph{Proceedings of the IEEE/ACM 48th International Conference on Software Engineering: Software Engineering in Practice}}. \bibinfo{pages}{212--223}.
\newblock
\href{https://doi.org/10.1145/3786583.3786866}{doi:\nolinkurl{10.1145/3786583.3786866}}


\bibitem[Geng et~al\mbox{.}(2026)]%
        {geng2025exploring}
\bibfield{author}{\bibinfo{person}{Francis Geng}, \bibinfo{person}{Anshul Shah}, \bibinfo{person}{Haolin Li}, \bibinfo{person}{Nawab Mulla}, \bibinfo{person}{Steve Swanson}, \bibinfo{person}{Gerald Soosai~Raj}, \bibinfo{person}{Daniel Zingaro}, {and} \bibinfo{person}{Leo Porter}.} \bibinfo{year}{2026}\natexlab{}.
\newblock \showarticletitle{Exploring student-AI interactions in vibe coding}. In \bibinfo{booktitle}{\emph{Proceedings of the 28th Australasian computing education conference}}. \bibinfo{pages}{45--54}.
\newblock
\href{https://doi.org/10.1145/3786228.3786236}{doi:\nolinkurl{10.1145/3786228.3786236}}


\bibitem[Goffman et~al\mbox{.}(1978)]%
        {goffman2023presentation}
\bibfield{author}{\bibinfo{person}{Erving Goffman} {et~al\mbox{.}}} \bibinfo{year}{1978}\natexlab{}.
\newblock \bibinfo{booktitle}{\emph{The presentation of self in everyday life}}. Vol.~\bibinfo{volume}{21}.
\newblock \bibinfo{publisher}{Harmondsworth London}.
\newblock


\bibitem[Guest et~al\mbox{.}(2006)]%
        {guest2006many}
\bibfield{author}{\bibinfo{person}{Greg Guest}, \bibinfo{person}{Arwen Bunce}, {and} \bibinfo{person}{Laura Johnson}.} \bibinfo{year}{2006}\natexlab{}.
\newblock \showarticletitle{How many interviews are enough? An experiment with data saturation and variability}.
\newblock \bibinfo{journal}{\emph{Field methods}} \bibinfo{volume}{18}, \bibinfo{number}{1} (\bibinfo{year}{2006}), \bibinfo{pages}{59--82}.
\newblock
\href{https://doi.org/10.1177/1525822X05279903}{doi:\nolinkurl{10.1177/1525822X05279903}}


\bibitem[Handa et~al\mbox{.}(2025)]%
        {handa2025economic}
\bibfield{author}{\bibinfo{person}{Kunal Handa}, \bibinfo{person}{Alex Tamkin}, \bibinfo{person}{Miles McCain}, \bibinfo{person}{Saffron Huang}, \bibinfo{person}{Esin Durmus}, \bibinfo{person}{Sarah Heck}, \bibinfo{person}{Jared Mueller}, \bibinfo{person}{Jerry Hong}, \bibinfo{person}{Stuart Ritchie}, \bibinfo{person}{Tim Belonax}, {et~al\mbox{.}}} \bibinfo{year}{2025}\natexlab{}.
\newblock \showarticletitle{Which economic tasks are performed with ai? evidence from millions of claude conversations}.
\newblock \bibinfo{journal}{\emph{arXiv preprint arXiv:2503.04761}} (\bibinfo{year}{2025}).
\newblock
\href{https://doi.org/10.48550/arXiv.2503.04761}{doi:\nolinkurl{10.48550/arXiv.2503.04761}}


\bibitem[Hao et~al\mbox{.}(2024)]%
        {hao2024empirical}
\bibfield{author}{\bibinfo{person}{Huizi Hao}, \bibinfo{person}{Kazi~Amit Hasan}, \bibinfo{person}{Hong Qin}, \bibinfo{person}{Marcos Macedo}, \bibinfo{person}{Yuan Tian}, \bibinfo{person}{Steven~HH Ding}, {and} \bibinfo{person}{Ahmed~E Hassan}.} \bibinfo{year}{2024}\natexlab{}.
\newblock \showarticletitle{An empirical study on developers’ shared conversations with ChatGPT in GitHub pull requests and issues}.
\newblock \bibinfo{journal}{\emph{Empirical Software Engineering}} \bibinfo{volume}{29}, \bibinfo{number}{6} (\bibinfo{year}{2024}), \bibinfo{pages}{150}.
\newblock
\href{https://doi.org/10.1007/s10664-024-10540-x}{doi:\nolinkurl{10.1007/s10664-024-10540-x}}


\bibitem[Humbatova et~al\mbox{.}(2020)]%
        {humbatova2020taxonomy}
\bibfield{author}{\bibinfo{person}{Nargiz Humbatova}, \bibinfo{person}{Gunel Jahangirova}, \bibinfo{person}{Gabriele Bavota}, \bibinfo{person}{Vincenzo Riccio}, \bibinfo{person}{Andrea Stocco}, {and} \bibinfo{person}{Paolo Tonella}.} \bibinfo{year}{2020}\natexlab{}.
\newblock \showarticletitle{Taxonomy of real faults in deep learning systems}. In \bibinfo{booktitle}{\emph{Proceedings of the ACM/IEEE 42nd international conference on software engineering}}. \bibinfo{pages}{1110--1121}.
\newblock
\href{https://doi.org/10.1145/3377811.3380395}{doi:\nolinkurl{10.1145/3377811.3380395}}


\bibitem[Jiang and Nam(2026)]%
        {jiang2025beyond}
\bibfield{author}{\bibinfo{person}{Shaokang Jiang} {and} \bibinfo{person}{Daye Nam}.} \bibinfo{year}{2026}\natexlab{}.
\newblock \showarticletitle{Beyond the prompt: An empirical study of cursor rules}. In \bibinfo{booktitle}{\emph{Proceedings of the 23rd International Conference on Mining Software Repositories}}. \bibinfo{pages}{397--409}.
\newblock
\href{https://doi.org/10.1145/3793302.3793367}{doi:\nolinkurl{10.1145/3793302.3793367}}


\bibitem[Jimenez et~al\mbox{.}(2024)]%
        {jimenez2023swe}
\bibfield{author}{\bibinfo{person}{Carlos~E Jimenez}, \bibinfo{person}{John Yang}, \bibinfo{person}{Alexander Wettig}, \bibinfo{person}{Shunyu Yao}, \bibinfo{person}{Kexin Pei}, \bibinfo{person}{Ofir Press}, {and} \bibinfo{person}{Karthik Narasimhan}.} \bibinfo{year}{2024}\natexlab{}.
\newblock \showarticletitle{Swe-bench: Can language models resolve real-world github issues?}. In \bibinfo{booktitle}{\emph{International Conference on Learning Representations}}, Vol.~\bibinfo{volume}{2024}. \bibinfo{pages}{54107--54157}.
\newblock


\bibitem[Kalliamvakou et~al\mbox{.}(2014)]%
        {kalliamvakou2014promises}
\bibfield{author}{\bibinfo{person}{Eirini Kalliamvakou}, \bibinfo{person}{Georgios Gousios}, \bibinfo{person}{Kelly Blincoe}, \bibinfo{person}{Leif Singer}, \bibinfo{person}{Daniel~M German}, {and} \bibinfo{person}{Daniela Damian}.} \bibinfo{year}{2014}\natexlab{}.
\newblock \showarticletitle{The promises and perils of mining github}. In \bibinfo{booktitle}{\emph{Proceedings of the 11th working conference on mining software repositories}}. \bibinfo{pages}{92--101}.
\newblock
\href{https://doi.org/10.1145/2597073.2597074}{doi:\nolinkurl{10.1145/2597073.2597074}}


\bibitem[Kaufman and Rousseeuw(1990)]%
        {kaufman2009finding}
\bibfield{author}{\bibinfo{person}{Leonard Kaufman} {and} \bibinfo{person}{Peter~J Rousseeuw}.} \bibinfo{year}{1990}\natexlab{}.
\newblock \bibinfo{booktitle}{\emph{Finding groups in data: an introduction to cluster analysis}}.
\newblock \bibinfo{publisher}{John Wiley \& Sons}.
\newblock
\href{https://doi.org/10.1002/9780470316801}{doi:\nolinkurl{10.1002/9780470316801}}


\bibitem[Kumar et~al\mbox{.}(2025)]%
        {kumar2025ai}
\bibfield{author}{\bibinfo{person}{Aayush Kumar}, \bibinfo{person}{Yasharth Bajpai}, \bibinfo{person}{Sumit Gulwani}, \bibinfo{person}{Gustavo Soares}, {and} \bibinfo{person}{Emerson Murphy-Hill}.} \bibinfo{year}{2025}\natexlab{}.
\newblock \showarticletitle{Why AI Agents Still Need You: Findings from Developer-Agent Collaborations in the Wild}. In \bibinfo{booktitle}{\emph{2025 40th IEEE/ACM International Conference on Automated Software Engineering (ASE)}}. IEEE, \bibinfo{pages}{432--444}.
\newblock
\href{https://doi.org/10.1109/ASE63991.2025.00043}{doi:\nolinkurl{10.1109/ASE63991.2025.00043}}


\bibitem[Landis and Koch(1977)]%
        {landis1977measurement}
\bibfield{author}{\bibinfo{person}{J~Richard Landis} {and} \bibinfo{person}{Gary~G Koch}.} \bibinfo{year}{1977}\natexlab{}.
\newblock \showarticletitle{The measurement of observer agreement for categorical data}.
\newblock \bibinfo{journal}{\emph{biometrics}} (\bibinfo{year}{1977}), \bibinfo{pages}{159--174}.
\newblock
\href{https://doi.org/10.2307/2529310}{doi:\nolinkurl{10.2307/2529310}}


\bibitem[Lazar et~al\mbox{.}(2017)]%
        {lazar2017research}
\bibfield{author}{\bibinfo{person}{Jonathan Lazar}, \bibinfo{person}{Jinjuan~Heidi Feng}, {and} \bibinfo{person}{Harry Hochheiser}.} \bibinfo{year}{2017}\natexlab{}.
\newblock \bibinfo{booktitle}{\emph{Research methods in human-computer interaction}}.
\newblock \bibinfo{publisher}{Morgan Kaufmann}.
\newblock


\bibitem[Lethbridge et~al\mbox{.}(2005)]%
        {lethbridge2005studying}
\bibfield{author}{\bibinfo{person}{Timothy~C Lethbridge}, \bibinfo{person}{Susan~Elliott Sim}, {and} \bibinfo{person}{Janice Singer}.} \bibinfo{year}{2005}\natexlab{}.
\newblock \showarticletitle{Studying Software Engineers: Data Collection Techniques for Software Field Studies: Lethbridge, Sim and Singer}.
\newblock \bibinfo{journal}{\emph{Empirical software engineering}} \bibinfo{volume}{10}, \bibinfo{number}{3} (\bibinfo{year}{2005}), \bibinfo{pages}{311--341}.
\newblock
\href{https://doi.org/10.1007/s10664-005-1290-x}{doi:\nolinkurl{10.1007/s10664-005-1290-x}}


\bibitem[Levenshtein et~al\mbox{.}(1966)]%
        {levenshtein1966binary}
\bibfield{author}{\bibinfo{person}{Vladimir~I Levenshtein} {et~al\mbox{.}}} \bibinfo{year}{1966}\natexlab{}.
\newblock \showarticletitle{Binary codes capable of correcting deletions, insertions, and reversals}. In \bibinfo{booktitle}{\emph{Soviet physics doklady}}, Vol.~\bibinfo{volume}{10}. Soviet Union, \bibinfo{pages}{707--710}.
\newblock


\bibitem[Li et~al\mbox{.}(2025b)]%
        {li2025rise}
\bibfield{author}{\bibinfo{person}{Hao Li}, \bibinfo{person}{Haoxiang Zhang}, {and} \bibinfo{person}{Ahmed~E Hassan}.} \bibinfo{year}{2025}\natexlab{b}.
\newblock \showarticletitle{The rise of ai teammates in software engineering (se) 3.0: How autonomous coding agents are reshaping software engineering}.
\newblock \bibinfo{journal}{\emph{arXiv preprint arXiv:2507.15003}} (\bibinfo{year}{2025}).
\newblock
\href{https://doi.org/10.48550/arXiv.2507.15003}{doi:\nolinkurl{10.48550/arXiv.2507.15003}}


\bibitem[Li et~al\mbox{.}(2025a)]%
        {li2025unveiling}
\bibfield{author}{\bibinfo{person}{Ruiyin Li}, \bibinfo{person}{Peng Liang}, \bibinfo{person}{Yifei Wang}, \bibinfo{person}{Yangxiao Cai}, \bibinfo{person}{Weisong Sun}, {and} \bibinfo{person}{Zengyang Li}.} \bibinfo{year}{2025}\natexlab{a}.
\newblock \showarticletitle{Unveiling the role of chatgpt in software development: Insights from developer-chatgpt interactions on github}.
\newblock \bibinfo{journal}{\emph{ACM Transactions on Software Engineering and Methodology}} (\bibinfo{year}{2025}).
\newblock
\href{https://doi.org/10.1145/3798163}{doi:\nolinkurl{10.1145/3798163}}


\bibitem[Liang et~al\mbox{.}(2024)]%
        {liang2024large}
\bibfield{author}{\bibinfo{person}{Jenny~T Liang}, \bibinfo{person}{Chenyang Yang}, {and} \bibinfo{person}{Brad~A Myers}.} \bibinfo{year}{2024}\natexlab{}.
\newblock \showarticletitle{A large-scale survey on the usability of ai programming assistants: Successes and challenges}. In \bibinfo{booktitle}{\emph{Proceedings of the 46th IEEE/ACM international conference on software engineering}}. \bibinfo{pages}{1--13}.
\newblock
\href{https://doi.org/10.1145/3597503.3608128}{doi:\nolinkurl{10.1145/3597503.3608128}}


\bibitem[Lui and Baldwin(2012)]%
        {lui2012langid}
\bibfield{author}{\bibinfo{person}{Marco Lui} {and} \bibinfo{person}{Timothy Baldwin}.} \bibinfo{year}{2012}\natexlab{}.
\newblock \showarticletitle{langid. py: An off-the-shelf language identification tool}. In \bibinfo{booktitle}{\emph{Proceedings of the ACL 2012 system demonstrations}}. \bibinfo{pages}{25--30}.
\newblock


\bibitem[Lyu et~al\mbox{.}(2025)]%
        {lyu2025my}
\bibfield{author}{\bibinfo{person}{Yunbo Lyu}, \bibinfo{person}{Zhou Yang}, \bibinfo{person}{Jieke Shi}, \bibinfo{person}{Jianming Chang}, \bibinfo{person}{Yue Liu}, {and} \bibinfo{person}{David Lo}.} \bibinfo{year}{2025}\natexlab{}.
\newblock \showarticletitle{"My productivity is boosted, but…" Demystifying Users’ Perception on AI Coding Assistants}. In \bibinfo{booktitle}{\emph{2025 40th IEEE/ACM International Conference on Automated Software Engineering (ASE)}}. IEEE, \bibinfo{pages}{191--203}.
\newblock
\href{https://doi.org/10.1109/ASE63991.2025.00024}{doi:\nolinkurl{10.1109/ASE63991.2025.00024}}


\bibitem[McDonald et~al\mbox{.}(2019)]%
        {mcdonald2019reliability}
\bibfield{author}{\bibinfo{person}{Nora McDonald}, \bibinfo{person}{Sarita Schoenebeck}, {and} \bibinfo{person}{Andrea Forte}.} \bibinfo{year}{2019}\natexlab{}.
\newblock \showarticletitle{Reliability and inter-rater reliability in qualitative research: Norms and guidelines for CSCW and HCI practice}.
\newblock \bibinfo{journal}{\emph{Proceedings of the ACM on human-computer interaction}} \bibinfo{volume}{3}, \bibinfo{number}{CSCW} (\bibinfo{year}{2019}), \bibinfo{pages}{1--23}.
\newblock
\href{https://doi.org/10.1145/3359174}{doi:\nolinkurl{10.1145/3359174}}


\bibitem[Mohsenimofidi et~al\mbox{.}(2026)]%
        {mohsenimofidi2025context}
\bibfield{author}{\bibinfo{person}{Seyedmoein Mohsenimofidi}, \bibinfo{person}{Matthias Galster}, \bibinfo{person}{Christoph Treude}, {and} \bibinfo{person}{Sebastian Baltes}.} \bibinfo{year}{2026}\natexlab{}.
\newblock \showarticletitle{Context engineering for AI agents in open-source software}. In \bibinfo{booktitle}{\emph{Proceedings of the 23rd International Conference on Mining Software Repositories}}. \bibinfo{pages}{194--198}.
\newblock
\href{https://doi.org/10.1145/3793302.3793350}{doi:\nolinkurl{10.1145/3793302.3793350}}


\bibitem[Mozannar et~al\mbox{.}(2024)]%
        {mozannar2024reading}
\bibfield{author}{\bibinfo{person}{Hussein Mozannar}, \bibinfo{person}{Gagan Bansal}, \bibinfo{person}{Adam Fourney}, {and} \bibinfo{person}{Eric Horvitz}.} \bibinfo{year}{2024}\natexlab{}.
\newblock \showarticletitle{Reading between the lines: Modeling user behavior and costs in AI-assisted programming}. In \bibinfo{booktitle}{\emph{Proceedings of the 2024 CHI conference on human factors in computing systems}}. \bibinfo{pages}{1--16}.
\newblock
\href{https://doi.org/10.1145/3613904.3641936}{doi:\nolinkurl{10.1145/3613904.3641936}}


\bibitem[Mysore et~al\mbox{.}(2025)]%
        {mysore2025prototypical}
\bibfield{author}{\bibinfo{person}{Sheshera Mysore}, \bibinfo{person}{Debarati Das}, \bibinfo{person}{Hancheng Cao}, {and} \bibinfo{person}{Bahareh Sarrafzadeh}.} \bibinfo{year}{2025}\natexlab{}.
\newblock \showarticletitle{Prototypical human-AI collaboration behaviors from LLM-Assisted writing in the wild}. In \bibinfo{booktitle}{\emph{Proceedings of the 2025 Conference on Empirical Methods in Natural Language Processing}}. \bibinfo{pages}{16830--16857}.
\newblock
\href{https://doi.org/10.18653/V1/2025.EMNLP-MAIN.852}{doi:\nolinkurl{10.18653/V1/2025.EMNLP-MAIN.852}}


\bibitem[Nam et~al\mbox{.}(2024)]%
        {nam2024using}
\bibfield{author}{\bibinfo{person}{Daye Nam}, \bibinfo{person}{Andrew Macvean}, \bibinfo{person}{Vincent Hellendoorn}, \bibinfo{person}{Bogdan Vasilescu}, {and} \bibinfo{person}{Brad Myers}.} \bibinfo{year}{2024}\natexlab{}.
\newblock \showarticletitle{Using an llm to help with code understanding}. In \bibinfo{booktitle}{\emph{Proceedings of the IEEE/ACM 46th International Conference on Software Engineering}}. \bibinfo{pages}{1--13}.
\newblock
\href{https://doi.org/10.1145/3597503.3639187}{doi:\nolinkurl{10.1145/3597503.3639187}}


\bibitem[Pimenova et~al\mbox{.}(2025)]%
        {pimenova2025good}
\bibfield{author}{\bibinfo{person}{Veronica Pimenova}, \bibinfo{person}{Sarah Fakhoury}, \bibinfo{person}{Christian Bird}, \bibinfo{person}{Margaret-Anne Storey}, {and} \bibinfo{person}{Madeline Endres}.} \bibinfo{year}{2025}\natexlab{}.
\newblock \showarticletitle{Good vibrations? A qualitative study of co-creation, communication, flow, and trust in vibe coding}.
\newblock \bibinfo{journal}{\emph{arXiv preprint arXiv:2509.12491}} (\bibinfo{year}{2025}).
\newblock
\href{https://doi.org/10.48550/arXiv.2509.12491}{doi:\nolinkurl{10.48550/arXiv.2509.12491}}


\bibitem[Rousseeuw(1987)]%
        {rousseeuw1987silhouettes}
\bibfield{author}{\bibinfo{person}{Peter~J Rousseeuw}.} \bibinfo{year}{1987}\natexlab{}.
\newblock \showarticletitle{Silhouettes: a graphical aid to the interpretation and validation of cluster analysis}.
\newblock \bibinfo{journal}{\emph{Journal of computational and applied mathematics}}  \bibinfo{volume}{20} (\bibinfo{year}{1987}), \bibinfo{pages}{53--65}.
\newblock
\href{https://doi.org/10.1016/0377-0427(87)90125-7}{doi:\nolinkurl{10.1016/0377-0427(87)90125-7}}


\bibitem[Sarkar and Drosos(2025)]%
        {sarkar2025vibe}
\bibfield{author}{\bibinfo{person}{Advait Sarkar} {and} \bibinfo{person}{Ian Drosos}.} \bibinfo{year}{2025}\natexlab{}.
\newblock \showarticletitle{Vibe coding: programming through conversation with artificial intelligence}. In \bibinfo{booktitle}{\emph{Proceedings of the 36th Annual Conference of the Psychology of Programming Interest Group (PPIG 2025)}}.
\newblock


\bibitem[Searle(1969)]%
        {searle1969speech}
\bibfield{author}{\bibinfo{person}{John~R Searle}.} \bibinfo{year}{1969}\natexlab{}.
\newblock \bibinfo{booktitle}{\emph{Speech acts: An essay in the philosophy of language}}.
\newblock \bibinfo{publisher}{Cambridge university press}.
\newblock
\href{https://doi.org/10.1017/CBO9781139173438}{doi:\nolinkurl{10.1017/CBO9781139173438}}


\bibitem[Sergeyuk et~al\mbox{.}(2026)]%
        {sergeyuk2026evolving}
\bibfield{author}{\bibinfo{person}{Agnia Sergeyuk}, \bibinfo{person}{Eric Huang}, \bibinfo{person}{Dariia Karaeva}, \bibinfo{person}{Anastasiia Serova}, \bibinfo{person}{Yaroslav Golubev}, {and} \bibinfo{person}{Iftekhar Ahmed}.} \bibinfo{year}{2026}\natexlab{}.
\newblock \showarticletitle{Evolving with AI: A Longitudinal Analysis of Developer Logs}.
\newblock \bibinfo{journal}{\emph{arXiv preprint arXiv:2601.10258}} (\bibinfo{year}{2026}).
\newblock
\href{https://doi.org/10.48550/arXiv.2601.10258}{doi:\nolinkurl{10.48550/arXiv.2601.10258}}


\bibitem[Shen and Tamkin(2026)]%
        {aiskillformation2026}
\bibfield{author}{\bibinfo{person}{Judy~Hanwen Shen} {and} \bibinfo{person}{Alex Tamkin}.} \bibinfo{year}{2026}\natexlab{}.
\newblock \showarticletitle{How AI impacts skill formation}.
\newblock \bibinfo{journal}{\emph{arXiv preprint arXiv:2601.20245}} (\bibinfo{year}{2026}).
\newblock
\href{https://doi.org/10.48550/arXiv.2601.20245}{doi:\nolinkurl{10.48550/arXiv.2601.20245}}


\bibitem[Siddiq et~al\mbox{.}(2024)]%
        {siddiq2024quality}
\bibfield{author}{\bibinfo{person}{Mohammed~Latif Siddiq}, \bibinfo{person}{Lindsay Roney}, \bibinfo{person}{Jiahao Zhang}, {and} \bibinfo{person}{Joanna Cecilia Da~Silva Santos}.} \bibinfo{year}{2024}\natexlab{}.
\newblock \showarticletitle{Quality assessment of chatgpt generated code and their use by developers}. In \bibinfo{booktitle}{\emph{Proceedings of the 21st international conference on mining software repositories}}. \bibinfo{pages}{152--156}.
\newblock
\href{https://doi.org/10.1145/3643991.3645071}{doi:\nolinkurl{10.1145/3643991.3645071}}


\bibitem[Sjoberg et~al\mbox{.}(2002)]%
        {sjoberg2002conducting}
\bibfield{author}{\bibinfo{person}{Dag~IK Sjoberg}, \bibinfo{person}{Bente Anda}, \bibinfo{person}{Erik Arisholm}, \bibinfo{person}{Tore Dyba}, \bibinfo{person}{Magne Jorgensen}, \bibinfo{person}{Amela Karahasanovic}, \bibinfo{person}{Espen~Frimann Koren}, {and} \bibinfo{person}{Marek Vok{\'a}c}.} \bibinfo{year}{2002}\natexlab{}.
\newblock \showarticletitle{Conducting realistic experiments in software engineering}. In \bibinfo{booktitle}{\emph{Proceedings international symposium on empirical software engineering}}. IEEE, \bibinfo{pages}{17--26}.
\newblock
\href{https://doi.org/10.1109/ISESE.2002.1166921}{doi:\nolinkurl{10.1109/ISESE.2002.1166921}}


\bibitem[Studer and Ritschard(2016)]%
        {studer2016matters}
\bibfield{author}{\bibinfo{person}{Matthias Studer} {and} \bibinfo{person}{Gilbert Ritschard}.} \bibinfo{year}{2016}\natexlab{}.
\newblock \showarticletitle{What matters in differences between life trajectories: A comparative review of sequence dissimilarity measures}.
\newblock \bibinfo{journal}{\emph{Journal of the Royal Statistical Society Series A: Statistics in Society}} \bibinfo{volume}{179}, \bibinfo{number}{2} (\bibinfo{year}{2016}), \bibinfo{pages}{481--511}.
\newblock
\href{https://doi.org/10.1111/rssa.12125}{doi:\nolinkurl{10.1111/rssa.12125}}


\bibitem[Tang et~al\mbox{.}(2024)]%
        {tang2024developer}
\bibfield{author}{\bibinfo{person}{Ningzhi Tang}, \bibinfo{person}{Meng Chen}, \bibinfo{person}{Zheng Ning}, \bibinfo{person}{Aakash Bansal}, \bibinfo{person}{Yu Huang}, \bibinfo{person}{Collin McMillan}, {and} \bibinfo{person}{Toby Jia-Jun Li}.} \bibinfo{year}{2024}\natexlab{}.
\newblock \showarticletitle{Developer behaviors in validating and repairing llm-generated code using ide and eye tracking}. In \bibinfo{booktitle}{\emph{2024 IEEE Symposium on Visual Languages and Human-Centric Computing (VL/HCC)}}. IEEE, \bibinfo{pages}{40--46}.
\newblock
\href{https://doi.org/10.1109/VL/HCC60511.2024.00015}{doi:\nolinkurl{10.1109/VL/HCC60511.2024.00015}}


\bibitem[Tang et~al\mbox{.}(2025a)]%
        {tang2025naturaledit}
\bibfield{author}{\bibinfo{person}{Ningzhi Tang}, \bibinfo{person}{David Meininger}, \bibinfo{person}{Gelei Xu}, \bibinfo{person}{Yiyu Shi}, \bibinfo{person}{Yu Huang}, \bibinfo{person}{Collin McMillan}, {and} \bibinfo{person}{Toby Jia-Jun Li}.} \bibinfo{year}{2025}\natexlab{a}.
\newblock \showarticletitle{NaturalEdit: Code Modification through Direct Interaction with Adaptive Natural Language Representation}.
\newblock \bibinfo{journal}{\emph{arXiv preprint arXiv:2510.04494}} (\bibinfo{year}{2025}).
\newblock
\href{https://doi.org/10.48550/arXiv.2510.04494}{doi:\nolinkurl{10.48550/arXiv.2510.04494}}


\bibitem[Tang et~al\mbox{.}(2025b)]%
        {tang2025exploring}
\bibfield{author}{\bibinfo{person}{Ningzhi Tang}, \bibinfo{person}{Emory Smith}, \bibinfo{person}{Yu Huang}, \bibinfo{person}{Collin McMillan}, {and} \bibinfo{person}{Toby Jia-Jun Li}.} \bibinfo{year}{2025}\natexlab{b}.
\newblock \showarticletitle{Exploring Direct Instruction and Summary-Mediated Prompting in LLM-Assisted Code Modification}. In \bibinfo{booktitle}{\emph{2025 IEEE Symposium on Visual Languages and Human-Centric Computing (VL/HCC)}}. IEEE, \bibinfo{pages}{68--80}.
\newblock
\href{https://doi.org/10.1109/VL-HCC65237.2025.00017}{doi:\nolinkurl{10.1109/VL-HCC65237.2025.00017}}


\bibitem[Timmermans and Tavory(2012)]%
        {timmermans2012theory}
\bibfield{author}{\bibinfo{person}{Stefan Timmermans} {and} \bibinfo{person}{Iddo Tavory}.} \bibinfo{year}{2012}\natexlab{}.
\newblock \showarticletitle{Theory construction in qualitative research: From grounded theory to abductive analysis}.
\newblock \bibinfo{journal}{\emph{Sociological theory}} \bibinfo{volume}{30}, \bibinfo{number}{3} (\bibinfo{year}{2012}), \bibinfo{pages}{167--186}.
\newblock
\href{https://doi.org/10.1177/0735275112457914}{doi:\nolinkurl{10.1177/0735275112457914}}


\bibitem[Van~der Maaten and Hinton(2008)]%
        {van2008visualizing}
\bibfield{author}{\bibinfo{person}{Laurens Van~der Maaten} {and} \bibinfo{person}{Geoffrey Hinton}.} \bibinfo{year}{2008}\natexlab{}.
\newblock \showarticletitle{Visualizing data using t-SNE.}
\newblock \bibinfo{journal}{\emph{Journal of machine learning research}} \bibinfo{volume}{9}, \bibinfo{number}{11} (\bibinfo{year}{2008}).
\newblock


\bibitem[Wang et~al\mbox{.}(2023)]%
        {wang2023large}
\bibfield{author}{\bibinfo{person}{Zhiqiang Wang}, \bibinfo{person}{Yiran Pang}, {and} \bibinfo{person}{Yanbin Lin}.} \bibinfo{year}{2023}\natexlab{}.
\newblock \showarticletitle{Large language models are zero-shot text classifiers}.
\newblock \bibinfo{journal}{\emph{arXiv preprint arXiv:2312.01044}} (\bibinfo{year}{2023}).
\newblock
\href{https://doi.org/10.48550/arXiv.2312.01044}{doi:\nolinkurl{10.48550/arXiv.2312.01044}}


\bibitem[Wei et~al\mbox{.}(2022)]%
        {wei2022chain}
\bibfield{author}{\bibinfo{person}{Jason Wei}, \bibinfo{person}{Xuezhi Wang}, \bibinfo{person}{Dale Schuurmans}, \bibinfo{person}{Maarten Bosma}, \bibinfo{person}{Fei Xia}, \bibinfo{person}{Ed Chi}, \bibinfo{person}{Quoc~V Le}, \bibinfo{person}{Denny Zhou}, {et~al\mbox{.}}} \bibinfo{year}{2022}\natexlab{}.
\newblock \showarticletitle{Chain-of-thought prompting elicits reasoning in large language models}.
\newblock \bibinfo{journal}{\emph{Advances in neural information processing systems}}  \bibinfo{volume}{35} (\bibinfo{year}{2022}), \bibinfo{pages}{24824--24837}.
\newblock
\href{https://doi.org/10.52202/068431-1800}{doi:\nolinkurl{10.52202/068431-1800}}


\bibitem[Welch(1947)]%
        {welch1947generalization}
\bibfield{author}{\bibinfo{person}{Bernard~L Welch}.} \bibinfo{year}{1947}\natexlab{}.
\newblock \showarticletitle{The generalization of ‘STUDENT'S’problem when several different population varlances are involved}.
\newblock \bibinfo{journal}{\emph{Biometrika}} \bibinfo{volume}{34}, \bibinfo{number}{1-2} (\bibinfo{year}{1947}), \bibinfo{pages}{28--35}.
\newblock
\href{https://doi.org/10.1093/biomet/34.1-2.28}{doi:\nolinkurl{10.1093/biomet/34.1-2.28}}


\bibitem[Xiao et~al\mbox{.}(2024)]%
        {xiao2024devgpt}
\bibfield{author}{\bibinfo{person}{Tao Xiao}, \bibinfo{person}{Christoph Treude}, \bibinfo{person}{Hideaki Hata}, {and} \bibinfo{person}{Kenichi Matsumoto}.} \bibinfo{year}{2024}\natexlab{}.
\newblock \showarticletitle{Devgpt: Studying developer-chatgpt conversations}. In \bibinfo{booktitle}{\emph{Proceedings of the 21st international conference on mining software repositories}}. \bibinfo{pages}{227--230}.
\newblock
\href{https://doi.org/10.1145/3643991.3648400}{doi:\nolinkurl{10.1145/3643991.3648400}}


\bibitem[Yang et~al\mbox{.}(2025)]%
        {yang2025spark}
\bibfield{author}{\bibinfo{person}{Yinuo Yang}, \bibinfo{person}{Ashley~Ge Zhang}, \bibinfo{person}{Steve Oney}, {and} \bibinfo{person}{April~Yi Wang}.} \bibinfo{year}{2025}\natexlab{}.
\newblock \showarticletitle{SPARK: Real-Time Monitoring of Multi-Faceted Programming Exercises}. In \bibinfo{booktitle}{\emph{2025 IEEE Symposium on Visual Languages and Human-Centric Computing (VL/HCC)}}. IEEE, \bibinfo{pages}{81--92}.
\newblock
\href{https://doi.org/10.1109/VL-HCC65237.2025.00018}{doi:\nolinkurl{10.1109/VL-HCC65237.2025.00018}}


\bibitem[Yang et~al\mbox{.}(2026)]%
        {yang2026lessons}
\bibfield{author}{\bibinfo{person}{Yinuo Yang}, \bibinfo{person}{Zheng Zhang}, \bibinfo{person}{Ningzhi Tang}, \bibinfo{person}{Xu Wang}, \bibinfo{person}{Alex Ambrose}, \bibinfo{person}{Nathaniel Myers}, \bibinfo{person}{Patrick Clauss}, {and} \bibinfo{person}{Toby Jia-Jun Li}.} \bibinfo{year}{2026}\natexlab{}.
\newblock \showarticletitle{Lessons from real-world deployment of a cognition-preserving writing tool: Students actively engage with critical thinking and planning affordances}.
\newblock \bibinfo{journal}{\emph{arXiv preprint arXiv:2603.15777}} (\bibinfo{year}{2026}).
\newblock
\href{https://doi.org/10.48550/arXiv.2603.15777}{doi:\nolinkurl{10.48550/arXiv.2603.15777}}


\bibitem[Zhao et~al\mbox{.}(2025)]%
        {zhao2025vibe}
\bibfield{author}{\bibinfo{person}{Songwen Zhao}, \bibinfo{person}{Danqing Wang}, \bibinfo{person}{Kexun Zhang}, \bibinfo{person}{Jiaxuan Luo}, \bibinfo{person}{Zhuo Li}, {and} \bibinfo{person}{Lei Li}.} \bibinfo{year}{2025}\natexlab{}.
\newblock \showarticletitle{Is vibe coding safe? Benchmarking vulnerability of agent-generated code in real-world tasks}.
\newblock \bibinfo{journal}{\emph{arXiv preprint arXiv:2512.03262}} (\bibinfo{year}{2025}).
\newblock
\href{https://doi.org/10.48550/arXiv.2512.03262}{doi:\nolinkurl{10.48550/arXiv.2512.03262}}


\bibitem[Zhao et~al\mbox{.}(2024)]%
        {zhao2024wildchat}
\bibfield{author}{\bibinfo{person}{Wenting Zhao}, \bibinfo{person}{Xiang Ren}, \bibinfo{person}{Jack Hessel}, \bibinfo{person}{Claire Cardie}, \bibinfo{person}{Yejin Choi}, {and} \bibinfo{person}{Yuntian Deng}.} \bibinfo{year}{2024}\natexlab{}.
\newblock \showarticletitle{Wildchat: 1m chatgpt interaction logs in the wild}. In \bibinfo{booktitle}{\emph{International Conference on Learning Representations}}, Vol.~\bibinfo{volume}{2024}. \bibinfo{pages}{34590--34605}.
\newblock


\bibitem[Zhong et~al\mbox{.}(2025a)]%
        {zhong2025vibe}
\bibfield{author}{\bibinfo{person}{Ming Zhong}, \bibinfo{person}{Xiang Zhou}, \bibinfo{person}{Ting-Yun Chang}, \bibinfo{person}{Qingze Wang}, \bibinfo{person}{Nan Xu}, \bibinfo{person}{Xiance Si}, \bibinfo{person}{Dan Garrette}, \bibinfo{person}{Shyam Upadhyay}, \bibinfo{person}{Jeremiah Liu}, \bibinfo{person}{Jiawei Han}, {et~al\mbox{.}}} \bibinfo{year}{2025}\natexlab{a}.
\newblock \showarticletitle{Vibe Checker: Aligning Code Evaluation with Human Preference}.
\newblock \bibinfo{journal}{\emph{arXiv preprint arXiv:2510.07315}} (\bibinfo{year}{2025}).
\newblock
\href{https://doi.org/10.48550/arXiv.2510.07315}{doi:\nolinkurl{10.48550/arXiv.2510.07315}}


\bibitem[Zhong et~al\mbox{.}(2025b)]%
        {zhong2025developer}
\bibfield{author}{\bibinfo{person}{Suzhen Zhong}, \bibinfo{person}{Ying Zou}, {and} \bibinfo{person}{Bram Adams}.} \bibinfo{year}{2025}\natexlab{b}.
\newblock \showarticletitle{Developer-llm conversations: An empirical study of interactions and generated code quality}.
\newblock \bibinfo{journal}{\emph{arXiv preprint arXiv:2509.10402}} (\bibinfo{year}{2025}).
\newblock
\href{https://doi.org/10.48550/arXiv.2509.10402}{doi:\nolinkurl{10.48550/arXiv.2509.10402}}


\end{thebibliography}

\end{document}